\documentclass[11pt]{article}

\usepackage[preprint]{acl}

\usepackage{times}
\usepackage{latexsym}

\usepackage[T1]{fontenc}
\usepackage[utf8]{inputenc}

\usepackage{microtype}

\usepackage{inconsolata}

\usepackage{graphicx}
\usepackage{amsmath,amsfonts} 
\usepackage{booktabs}
\usepackage{algorithmic}
\usepackage{algorithm}
\usepackage{array}
\usepackage{url}
\title{Brain-Inspired Exploration of Functional Networks and Key Neurons in Large Language Models}

\author{
 \textbf{Yiheng Liu\textsuperscript{1}},
 \textbf{Zhengliang Liu\textsuperscript{2}},
 \textbf{Zihao Wu\textsuperscript{2}},
 \textbf{Junhao Ning\textsuperscript{1}},
 \textbf{Haiyang Sun\textsuperscript{1}},
 \textbf{Sichen Xia\textsuperscript{1}},
 \textbf{Yang Yang\textsuperscript{1}},
 \\
 \textbf{Xiaohui Gao \textsuperscript{1}},
 \textbf{Ning Qiang\textsuperscript{3}},
 \textbf{Bao Ge\textsuperscript{3}},
 \textbf{Tianming Liu\textsuperscript{2}},
 \textbf{Junwei Han\textsuperscript{1}},
 \textbf{Xintao Hu\textsuperscript{1}},
\\
 \textsuperscript{1}School of Automation, Northwestern Polytechnical University, Xi'an, China
 \\
 \textsuperscript{2}School of Computing, University of Georgia, Athens, USA
 \\
 \textsuperscript{3}School of Physics and Information Technology, Shaanxi Normal University, Xi'an, China
  \\
 \small{
   \textbf{Correspondence:} \href{xhu@nwpu.edu.cn}{xhu@nwpu.edu.cn}
 }
}

\begin{document}
\maketitle
\begin{abstract}
In recent years, the rapid advancement of large language models (LLMs) in natural language processing has sparked significant interest among researchers to understand their mechanisms and functional characteristics. Although prior studies have attempted to explain LLM functionalities by identifying and interpreting specific neurons, these efforts mostly focus on individual neuron contributions, neglecting the fact that human brain functions are realized through intricate interaction networks. Inspired by research on functional brain networks (FBNs) in the field of neuroscience, we utilize similar methodologies estabilished in FBN analysis to explore the "functional networks" within LLMs in this study. Experimental results highlight that, much like the human brain, LLMs exhibit certain functional networks that recur frequently during their operation. Further investigation reveals that these functional networks are indispensable for LLM performance. Inhibiting key functional networks severely impairs the model's capabilities. Conversely, amplifying the activity of neurons within these networks can enhance either the model's overall performance or its performance on specific tasks. This suggests that these functional networks are strongly associated with either specific tasks or the overall performance of the LLM. Code is available at \url{https://github.com/WhatAboutMyStar/LLM_ACTIVATION}.
\end{abstract}

\section{Introduction}
In recent years, large language models (LLMs) have become a focal point of research in the field of artificial intelligence (AI) due to their remarkable capabilities in natural language processing \cite{zhao2024explainability,zhao2023survey,liu2023summary,WANG2024,LIU2025129190}. However, these models are often considered "black boxes", with insufficient understanding of their internal mechanisms. Establishing methods to explain and understand LLMs is essential both for improving model transparency and trustworthiness and for establishing a foundation to develop more efficient and reliable AI systems. 

One research direction on the mechanistic interpretability of LLMs focuses on the functional role of individual neurons \cite{yu2024interpreting,dai-etal-2022-knowledge,yu2024neuron,niu2024what,chen2024journey}. Prior studies have shown the specific functional roles of LLM neurons \cite{alkhamissi2024llm,wang-wen-etal2022skill}. For example, some neurons may specialize in processing linguistic structures, while others might be responsible for reasoning tasks \cite{huo-etal-2024-mmneuron,zhao2024how}. Removal of certain neurons leads to a significant degradation in model performance. In addition, by manipulating certain neurons (for example, amplifying their output signals \cite{song2024does,duan2025unveiling}), it is possible to enhance an LLM’s performance on specific tasks. This suggests that some neurons are strongly associated with particular tasks.

Methods for identifying key neurons within LLM can be categorized into several categories. These include analyzing the gradients of neurons to evaluate their impact on model predictions \cite{sundararajan2017axiomatic,lundstrom2022rigorous}, employing causal tracing techniques to uncover the causal relationships that influence model behavior \cite{nikankin2024arithmetic}, and conducting statistical analyzes of activated neurons to measure their information content and variability \cite{alkhamissi2024llm,song2024does,tang2024language}. These approaches provide valuable tools for understanding and explaining LLMs, offering deeper insight into their inner mechanisms.

However, the function of an individual neuron in human brain is much more complex than it might initially seem. Neurons in human brain often form functional networks through their interactions and connectivity, collaboratively working to perform higher-level cognitive tasks \cite{smith2009correspondence,bullmore2009complex}. The role of a neuron therefore extends beyond its individual activation patterns and is shaped by its cooperation with other neurons within these networks \cite{bullmore2009complex,liu2024spatial,liu2024mapping}. In this context, current key neuron identification studies largely overlook the functional network perspective and fail to consider the coordinated roles of neurons. As a result, these limitations have hindered a deeper understanding of neuronal function, neglecting the insights offered by neuroscience research on functional brain networks (FBNs) \cite{hassabis2017neuroscience,vilasposition}. 

In this study, we draw inspiration from neuroscience to investigate whether LLMs contain functional networks similar to those found in the human brain. By recognizing the similarities between functional magnetic resonance imaging (fMRI) \cite{matthews2004functional,logothetis2008we} signals and the output signals of neurons in LLM, we hypothesize that the techniques used in fMRI analysis could be adapted to analyze LLM neurons. Specifically, we treat the neuron outputs from the multilayer perception (MLP) layers of LLMs as analogous to fMRI signals and applied Independent Component Analysis (ICA) \cite{hyvarinen2000independent,beckmann2005investigations,varoquaux2010group} to decompose these neuron outputs into multiple functional networks.

Our experiments on extensive datasets confirmed the existence of functional networks within LLMs, conceptually mimicking deriving FBNs from fMRI data \cite{mensch2016compressed,varoquaux2010ica,liu2023spatial,he2023multi,ge2020discovering,lv2015sparse}. Some functional networks exhibit highly consistent spatial organization across diverse inputs and play a critical role in model functionality. Inhibition of specific key networks (typically comprising fewer than 2\% of the model’s neurons) significantly degrades performance, while amplifying these critical networks can improve the performance of the model in a specific task or overall.

Our contributions are summarized as follows:

1. Inspired by functional brain network analysis in neuroscience, we have introduced an analytical framework based on Independent Component Analysis (ICA) to explore "functional networks" within LLMs. This approach moves beyond treating activations as undifferentiated representations and instead reveals structured, distributed subnetworks that consistently co-activate across inputs.

2. We have demonstrated that LLMs exhibit functional networks, which share a notable similarity with the human brain in that both demonstrate stable functional patterns.

3. We have demonstrated that neurons within these functional networks are strongly associated with specific tasks and are essential to maintain the functionality of LLMs. 

\begin{figure*}[ht]
    \begin{center}
    \centerline{\includegraphics[width=\textwidth]{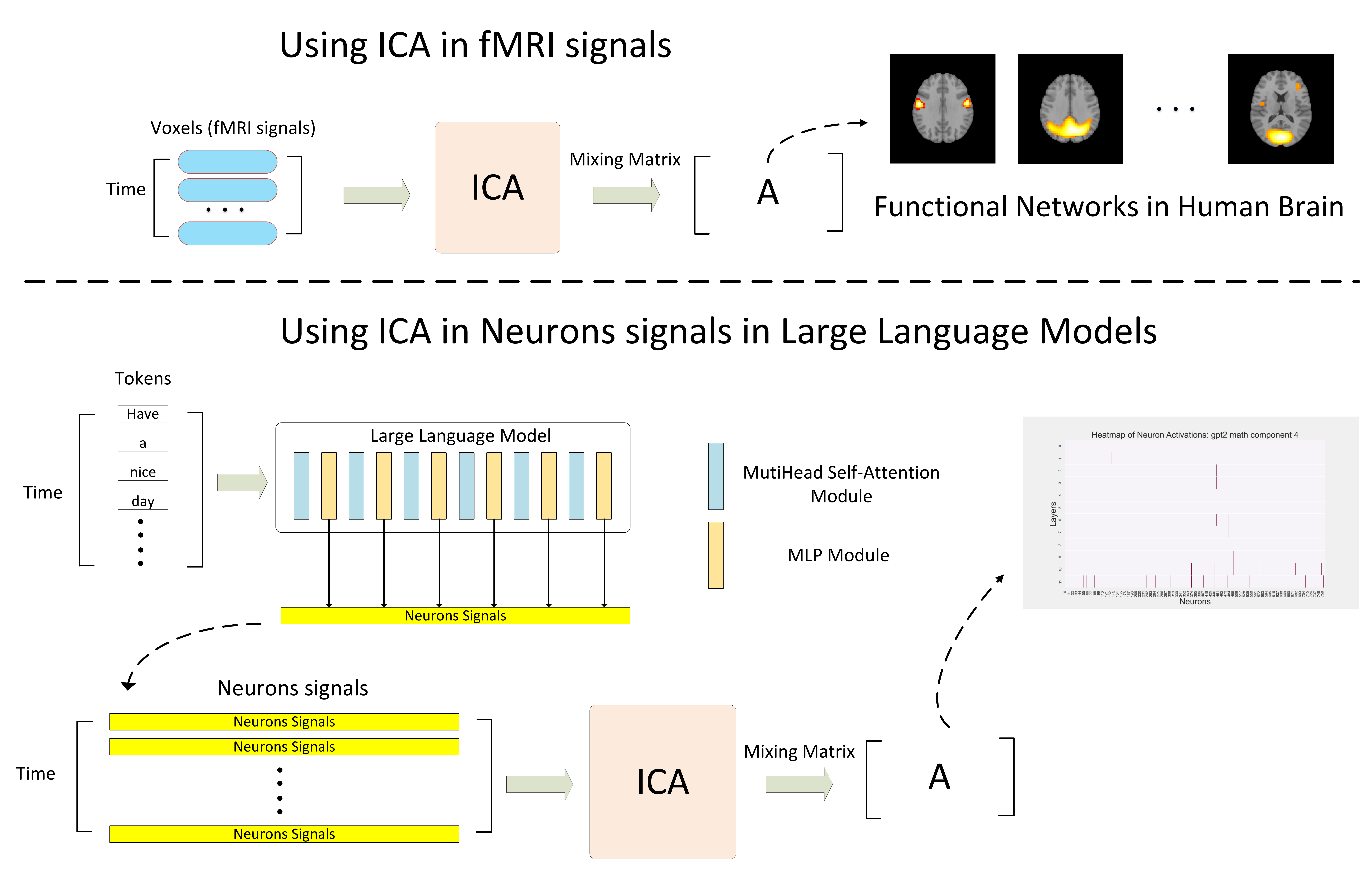}}
    \caption{The framework of brain inspired exploration of functional networks within LLMs. (a) Identifying functional brain networks from whole-brain fMRI signals via ICA. (b) Identifying functional networks from responses of artificial neurons within LLMs.}
    \label{fig::frame}
    \end{center}
\end{figure*}

\section{Preliminaries}

\textbf{Transformer, MLP layer:} LLMs utilized in this study are based on the transformer architecture \cite{vaswani2017attention}, specifically employing a decoder-only configuration \cite{GPT,GPT2,GPT3,yang2024qwen2,glm2024chatglm}. In this configuration, each transformer decoder consists of two primary components: a multi-head self-attention module and an MLP module. Typically, an MLP module consists of two fully connected layers. The first layer increases the dimensionality, often to four times the original dimension, followed by a non-linear activation function. The second layer then reduces the dimensionality back to its original size. In our study, we focus on the neurons located in the final MLP layer of each decoder module within the model. Given an input vector $\mathbf{x} \in \mathbb{R}^{d_{\text{model}}}$, the MLP module can be represented as follows:

\begin{equation}
    \text{MLP}(\mathbf{x}) = \mathbf{W}_2 \cdot \sigma (\mathbf{W}_1 \cdot \mathbf{x} + \mathbf{b}_1) + \mathbf{b}_2
\end{equation}

$\mathbf{W}$ is the weight matrix of the linear transformation, $\mathbf{b}$ is the bias vector of the linear transformation, $\sigma$ is a non-linear activation function. The neuron outputs used in this paper are the outputs of the $\text{MLP}(\mathbf{x})$. The output of neurons in the MLP layer is also the output of the last component of the Transformer block, which can better reflect the information processing results of the entire module, possessing stronger integrity and representativeness.

\textbf{Functional Brain Networks, FBNs:} FBNs refer to collections of brain regions that are co-activated during specific tasks or while at rest \cite{dong2020discovering}. FMRI is a non-invasive technique used to measure Blood-Oxygen Level Dependent (BOLD) signals, which reflect neuronal activity indirectly \cite{matthews2004functional,logothetis2008we}. The intensity of voxel values in fMRI signals indirectly reflects neuronal activity by capturing variations in local blood oxygen levels due to neural metabolism. Neuroscientific research hypothesizes that the observed BOLD signals in fMRI signals are likely the result of multiple independent functional networks working together. Essentially, these BOLD signals can be considered as linear combinations of several source signals, each representing a distinct functional network. 

\textbf{Independent Component Analysis, ICA:} ICA is a powerful data-driven technique used to extract source signals that are as statistically independent as possible from a mixed signal. In the field of neuroscience, ICA is widely applied to fMRI data to uncover underlying FBNs \cite{hyvarinen2000independent}. It disentangles mixed fMRI signals into several independent components, where each component represents a distinct functional network \cite{varoquaux2010group}. Each extracted independent component is associated with a spatial map that illustrates which brain regions contribute to that component. These regions typically display synchronized activities, indicating their coordinated involvement in specific neural processes.

The objective of ICA is to recover the source signals $\mathbf{S}$ from the observed signals $\mathbf{X}$. Suppose that we have $n$ observed signals $\left[ \mathbf{x}_1, \mathbf{x}_2, \ldots, \mathbf{x}_n \right]$, which are linear mixtures of $m$ independent source signals $\left[ \mathbf{s}_1, \mathbf{s}_2, \ldots, \mathbf{s}_m \right]$. The relationship between the observed signals $\mathbf{X}$ and the source signals $\mathbf{S}$ can be expressed as:
\begin{equation}
    \mathbf{X} = \mathbf{A} \mathbf{S}
\end{equation}
where $\mathbf{A}$ is the mixing matrix that describes how the source signals are combined to produce the observed signals. Each row of the linear mixing matrix represents the spatial pattern of the corresponding functional network. In this study, the functional networks derived from LLMs neuron signals refer to the rows of the linear mixing matrix $\mathbf{A}$, which indicate the set of neurons that are consistently co-activated under different conditions. %We adopt FastICA \cite{hyvarinen2000independent}, an efficient algorithm for implementing ICA, in this study to derive FBNs within LLMs. 

\section{Method}
\subsection{Datasets and Models}
Five pre-trained LLM models, including GPT2 \cite{GPT2}, Qwen2.5-3B-Instruct \cite{qwen2}, Qwen2.5-7B-Instruct \cite{qwen2}, ChatGLM3-6B-base \cite{glm2024chatglm} were used to validate the proposed method. The consistency and variations of the functional networks within these LLM models when processing various external stimuli were assessed by exposing them to text samples from each of four datasets, including the AGNEWS dataset \cite{zhang2015character}, encyclopedia entries from Wikitext2 \cite{merity2016pointer}, mathematical texts from MathQA \cite{amini2019mathqa}, and code snippets from the CodeNet dataset \cite{codenet}. In addition, some benchmark datasets including SQUAD \cite{rajpurkar-etal-2016-squad}, GLUE \cite{wang2018glue}, and AGNEWS \cite{zhang2015character} were used to test the performance of the LLM models after editing the key functional networks identified in LLMs.

\subsection{Artificial Neurons and Neuron Signals within LLMs}
There is a certain similarity between fMRI signals and the neuron signals in LLMs. fMRI signals reflect the activity of biological neurons in the brain when subjected to external stimuli or under resting state, while the neuron signals of artificial neurons in LLMs represent the internal state changes of the model as it processes input information. Both fMRI signals and the neuron signals in LLMs exhibit temporal characteristics. In fMRI, different stimuli trigger activity changes in different regions of the brain at different time points. Similarly, in LLMs, different inputs causes variations in neuron output at different time steps. This temporal characteristic allows us to analyze these signals dynamically, looking for coordinated activities and regularities within them.

To extract functional networks from LLMs, the first step is to define artificial neurons (ANs) in LLMs, analogous to voxels in fMRI data. In this study, we utilize the neurons in the last layer of the MLP module in each Transformer blocks (Fig. ~\ref{fig::frame}). Afterwards, the neuron signal of an AN, analogous to the fMRI signal of a voxel, is collected as the responses of the AN subjected to text inputs. 

\subsection{Identifying Functional Networks within LLMs via ICA}
Both fMRI signals and neuron signals in LLMs can be viewed as mixtures of multiple independent components. In the brain, different neural activity patterns combine to form complex fMRI signals. In LLMs, interactions among different functional modules result in the output signals being a superposition of multiple independent components. ICA is specifically designed for such signal-mixing scenarios and can effectively separate these independent components. In fMRI analysis, the independent components separated by ICA correspond to different functional networks in the brain (Fig.~\ref{fig::frame}(a)), such as the visual network or auditory network. Likewise, in LLMs, the independent components separated by ICA may correspond to different functional modules within the model (Fig.~\ref{fig::frame}(b)). Therefore, ICA can help reveal the functional organization inside LLMs.

\subsection{Validation and Evaluation}
\label{sec::validation}
\subsubsection{Consistency of Functional Networks within LLMs}

Identifying FBNs in fMRI data typically involves performing ICA on individual-level and group-level. In individual level analysis, ICA is applied to single-subject fMRI data. In contrast, in group level analysis, ICA is performed on data from multiple subjects to derive common brain networks shared by multiple subjects. Individual-level ICA captures subject-specific functional networks that are either spontaneous or task-evoked, offering high individuality but suffering from unstable component number and interpretation across subjects. Group-level ICA, in contrast, identifies reproducible, shared functional networks across subjects, providing greater stability and comparability. We use group-level ICA to extract stable, interpretable, and comparable functional network templates. 

Following the similar strategy, we randomly selected 100 samples from each dataset, analogous to select 100 subjects in group-level fMRI analysis. We then randomly selected another 100 samples from the same dataset for individual-level analysis. By comparing functional networks derived from individual-level and group-level ICA, we investigate whether similarly stable "functional networks" also exist in LLMs.

The functional networks derived from the group-level ICA analysis can be considered as templates representing a set of common functional networks. We then assessed the spatial similarity between these templates and the functional networks derived from the individual-level analysis to evaluate the consistency and variations of functional networks in LLMs. Here, we adopted intersection over union (IoU), , which is commonly used in neuroscience studies, as a metric to measure the spatial similarity between two functional networks $N^{(1)}$ and $N^{(2)}$. The IoU is defined as follows:

\begin{equation}
    IoU(N^{(1)}, N^{(2)}) = \frac{\sum_{i=1}^{n}|N_{i}^{(1)} \cap N_{i}^{(2)}|}{\sum_{i=1}^{n}|N_{i}^{(1)} \cup N_{i}^{(2)}|}
\end{equation}
where $n$ represents the number of neurons in the corresponding functional network. It is worth noting that, similar to comparing functional networks in neuroscience, calculating spatial similarities between functional networks in LLMs also requires applying a threshold to the mixing matrix $\mathbf{A}$ to filter out neurons with lower activation values. We set the threshold as 3.6, a value commonly used in functional brain network analysis.

According to our experimental findings and experience in functional brain network analysis, two networks appear reasonably similar when the IoU between them exceeds 0.2. Consequently, such networks are usually classified into the same category of functional networks. In this study, a template of LLM functional network extracted in group level analysis is considered consistent if its counterpart (IoU greater than 0.2) in individual level analysis can be reliably identified, representing a consistent functional network within an LLM.   

\subsubsection{Functional Networks Editing}
We conducted functional network editing experiments to investigate how targeted manipulation of neurons influences model behavior. Specifically, we first identified functional networks using ICA, then directly edited the corresponding neurons during inference, either by inhibiting them (setting their activations to zero) or amplifying them (scaling up their outputs). First, we edit these functional networks and applied on the same task and dataset to assess their task-specific role. Second, we applied the same neuron edits across different tasks and datasets to evaluate whether these functional networks support general, cross-task capabilities.

\section{Results}
\subsection{Consistent Functional Networks within LLMs}
\label{sec::consistent}
In the experiment, we derived 10 functional network templates in group-level analysis and 64 functional networks in each of the 100 individual-level analysis. For each group level template, we counted the number of its counterparts (see ~\ref{sec::validation} for details) in all 100 individual-level analysis, as summarized in Tables ~\ref{tab::gpt_cnt}, ~\ref{tab::qwen_cnt}, ~\ref{tab::qwen7b_cnt}, ~\ref{tab::glm_cnt} for GPT-2, Qwen2.5-3B-Instruct, Qwen2.5-7B-Instruct and ChatGLM3-6B-base, respectively (We include the Qwen and GLM results in the appendix.). 

\begin{table}[t]
    \caption{The number of the counterparts in 100 individual-level analysis for the functional network templates in group-level analysis in GPT-2.}
    \label{tab::gpt_cnt}
    \begin{center}
    \resizebox{\columnwidth}{!}{
    \begin{small}
    \begin{sc}
        \begin{tabular}{ccccc}
            \toprule
            Templates & News & Wiki & Math & code \\
            \midrule
            1  & 139  & 1848 & 219  & 1151 \\
            2  & 236  & 1830 & 165  & 1198 \\
            3  & 243  & 47   & 5    & 59   \\
            4  & 316  & 1722 & 116  & 1230 \\
            5  & 307  & 1301 & 74   & 1209 \\
            6  & 416  & 120  & 54   & 1352 \\
            7  & 16   & 1080 & 88   & 1473 \\
            8  & 107  & 529  & 114  & 959  \\
            9  & 92   & 37   & 125  & 915  \\
            10 & 330  & 450  & 190  & 975  \\
            \bottomrule
        \end{tabular}
    \end{sc}
    \end{small}
    }
    \end{center}
\end{table}

We observed that, across all tested LLMs, most group-level functional network templates have corresponding counterparts in individual-level analysis. Given that 100 individual models were analyzed, each with 64 functional networks (totaling 6,400 per dataset), the widespread presence of group-level templates in individual analyses indicates strong reproducibility. In GPT-2 (Table~\ref{tab::gpt_cnt}), match counts are exceptionally high—some templates appear thousands of times (e.g., Template 7 in Code: 1,473 matches). This reflects both the robustness of the group-level templates and the relative functional simplicity of smaller models, where fewer, more dominant networks emerge consistently across individuals.

In contrast, for larger models like Qwen2.5-3B-Instruct (Table~\ref{tab::qwen_cnt}), Qwen2.5-7B-Instruct (Table~\ref{tab::qwen7b_cnt}), and GLM (Tables~\ref{tab::glm_cnt}), the number of matches per template is generally lower, but this is not due to instability. Instead, as model scale increases, the total number of distinct functional networks grows, leading to greater functional diversity and specialization. Consequently, individual networks are more distributed across a larger repertoire, diluting the frequency of any single template. Nevertheless, key templates still emerge consistently across individuals, such as Template 7 in Qwen2.5-3B (Wiki: 89; Code: 62) and Template 10 in Qwen2.5-7B (Code: 119; Math: 191), demonstrating that stable, shared functional organizations persist even in larger models.

We next evaluate whether the same functional networks emerge across different tasks in LLMs, aiming to identify shared, task-invariant functional structures. Table~\ref{tab::qwen-7b-template} shows the spatial similarity between group-level functional network templates in Qwen2.5-7B-Instruct across News, Wiki, Math and Code tasks.

The results reveal strong cross-task consistency in specific functional networks. These high IoU values indicate that certain functional networks are not task-specific but instead reappear across diverse domains. This parallels findings in human neuroscience, where core brain networks like the default mode network \cite{raichle2007default,raichle2015brain} activate across a wide range of cognitive tasks. 

\begin{table}[t]
\caption{Spatial similarity between functional network templates in group-level analysis in Qwen2.5-7B-Instruct.}
\label{tab::qwen-7b-template}
\begin{center}
\resizebox{\columnwidth}{!}{
\begin{small}
\begin{sc}
\begin{tabular}{ccc}
\toprule
Template 1 & Template 2 & IoU \\
\midrule
News Component 3 & Code Component 2 & 0.8605 \\
News Component 6 & Code Component 3 & 0.7018 \\
News Component 7 & Code Component 2 & 0.7959 \\
Code Component 9 & Wiki Component 7 & 0.8837 \\
Code Component 2 & Math Component 1 & 0.8723 \\
Code Component 2 & Math Component 2 & 0.9091 \\
Code Component 2 & Math Component 3 & 0.8043 \\
Code Component 2 & Math Component 5 & 0.9070 \\
Code Component 9 & Math Component 10 & 0.9091 \\
News Component 6 & Wiki Component 7 & 0.7255 \\
News Component 3 & Math Component 1 & 0.7872 \\
News Component 3 & Math Component 2 & 0.8605 \\
News Component 3 & Math Component 3 & 0.8372 \\
News Component 3 & Math Component 5 & 0.9024 \\
News Component 6 & Math Component 10 & 0.6852 \\
Wiki Component 7 & Math Component 10 & 0.9302 \\
\bottomrule
\end{tabular}
\end{sc}
\end{small}
}
\end{center}
\end{table}

\subsection{Functional Networks Editing Experiments}

\subsubsection{Inhibiting Functional Networks}

We first adopt the neuron-lesion experiment, we selectively deactivate functional networks within the LLMs to assess the effects of functional network removal on overall performance and functional behavior. 

For Qwen2.5 and ChatGLM, we used lm eval framework and carefully crafted prompts to evaluate their performance in a zero-shot setting. In our assessments, we used accuracy, F1-score, Matthews Correlation Coefficient (MCC) and word perplexity as the performance metric.

\begin{table}[t]
\caption{Performance of Qwen2.5-7B-Instruct with inhibited functional networks on the SST-2 Dataset. The first column represents each functional network and the number of neurons associated with it. The second column to forth column shows the accuracy in SST-2, CoLA and MRPC. These functional networks derived from the SST-2 dataset. The model's accuracy under normal conditions (without lesion) is 0.9358 (SST-2), 0.6913 (CoLA), 0.6765 (MRPC) and 0.8448 (RTE).}
\label{tab::qwen7b_10_sst2}
\begin{center}
\resizebox{\columnwidth}{!}{
\begin{small}
\begin{sc}
\begin{tabular}{ccccc}
\toprule
Networks & \multicolumn{4}{c}{\textbf{Accuracy}} \\
(100352) & SST-2 &  CoLA & MRPC & RTE\\
\midrule
1 $\rightarrow$ 41 & 0.0963 & 0.1074 & 0.0221 & 0.0072\\
2 $\rightarrow$ 336 & 0.8085 & 0.6913 & 0.3799 & 0.5271\\
3 $\rightarrow$ 356 & 0.8131 & 0.6913 & 0.5907 & 0.5271\\
4 $\rightarrow$ 43 & 0.8417 & 0.6913 &  0.5294  & 0.5271\\
5 $\rightarrow$ 239 & 0.8016 & 0.6913 & 0.5196  & 0.4982\\
6 $\rightarrow$ 40 & 0.0401 & 0.0662 &  0.1838  & 0.0036 \\
7 $\rightarrow$ 41 & 0.1456 & 0.0518 &  0.2672  & 0.0361  \\
8 $\rightarrow$ 41 & 0.1892 & 0.1850 &  0.2059  & 0.0144   \\
9 $\rightarrow$ 53 & 0.8360 & 0.6913 &  0.3922  & 0.5271    \\
10 $\rightarrow$ 45 & 0.8314 & 0.6913 & 0.3333  &  0.5271    \\
\bottomrule
\end{tabular}
\end{sc}
\end{small}
}
\end{center}
\end{table}

Table~\ref{tab::qwen7b_10_sst2} presents the performance results of the Qwen2.5-7B-Instruct model on the SST-2 dataset after selectively inhibiting specific functional networks derived in group-level analysis. The results indicate that inhibiting these functional networks leads to varying degrees of performance degradation. Notably, inhibiting just a few dozens of key neurons can significantly impair the model's performance. In contrast, random neuron inhibiting has a negligible impact on model performance, as shown in Table~\ref{tab::qwen7b_10_all} (see in the appendix). It is seen that even when up to 15\% neurons are randomly inhibiting, the performance drop remains minimal. This finding is consistent with previous studies that have also observed that random neuron inhibiting does not substantially affect model performance \cite{alkhamissi2024llm,song2024does}. Taken together, our experimental results emphasize the critical role of specific functional networks identified through ICA. These networks appear to encode essential information that is vital for the model's capability. 

We also observed that some functional networks identified on the SST-2 dataset are not only crucial for SST-2 but also play an important role in other tasks. Inhibiting these networks not only degrades the model's performance on SST-2 but also leads to a decline in performance on other tasks. In contrast, networks that are less important for SST-2 may still be critical for other tasks. This observation suggests that the model may contain shared underlying structures or patterns that are not task-specific but instead capture general linguistic features or principles. 

\begin{figure}[ht]
    \begin{center}
    \centerline{\includegraphics[width=\columnwidth]{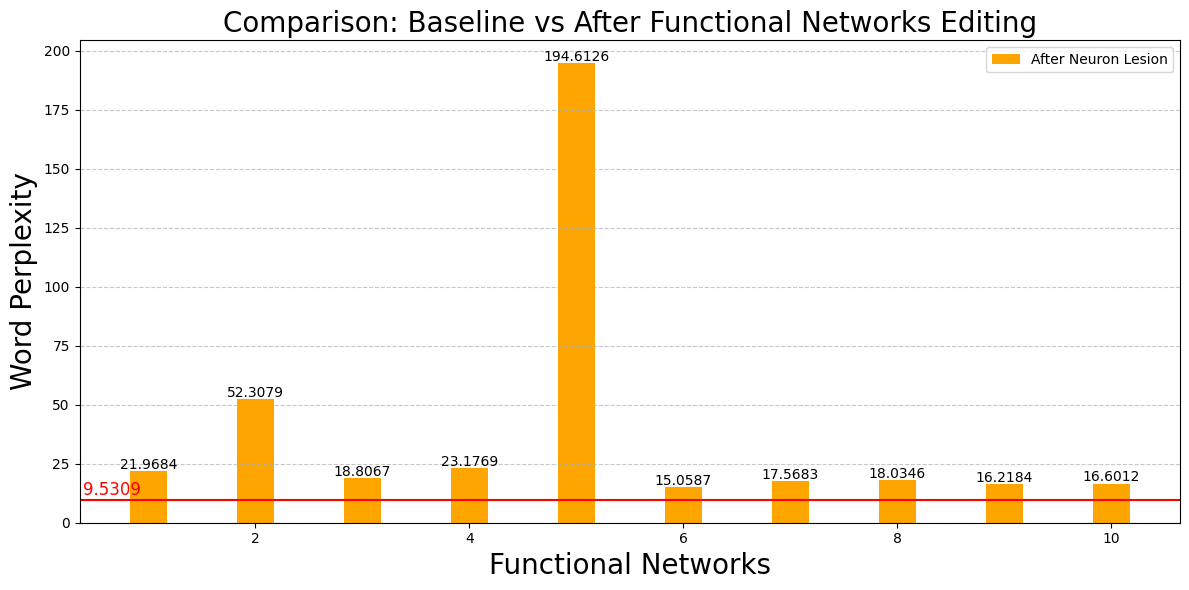}}
    \caption{Perplexity results for Qwen2.5-7B-Instruct after inhibiting the neurons corresponding to each of 10 functional networks obtained from SST-2 dataset, respectively.}
    \label{fig::qwen_ppl}
    \end{center}
\end{figure}

Table~\ref{tab::qwen7b_10_all} and Table~\ref{tab::glm_10_all} (see in the appendix) present the performance results of Qwen2.5-7B-Instruct and ChatGLM3-6B-base by inhibiting neurons within specific functional networks. These models rely on well-designed prompts to zero-shot generate appropriate text for various tasks. When functional networks are inhibited, the models' ability to generate coherent and task-relevant text is severely compromised. 

In addition to evaluating zero-shot classification performance, we also assessed the language modeling capability of LLMs with inhibited functional networks using the Wikitext dataset, measured by word-level perplexity. 

We conducted experiments on the Qwen2.5-7B-Instruct model, using the same 10 functional network templates previously derived from the SST-2 dataset. The results are shown in the Figure~\ref{fig::qwen_ppl}. After inhibiting different functional networks, the model's language modeling performance degraded to varying degrees, with functional network templates 2 and 5 causing the largest increases in perplexity.

\subsubsection{Amplifying Functional Networks}
To further investigate the functional roles of key neurons within these functional networks, we performed functional networks editing experiments by amplifying their activations. Specifically, we tested both additive and multiplicative editing strategies and found that multiplicative editing yielded the most stable and consistent effects. Our experiments show that scaling neuron activations by a factor of 1.02 yields the best results. Therefore, all experiments presented in this section use 1.02 as the multiplicative editing factor. Results of ablation studies on additive editing and other multiplicative factors can be found in the appendix.

\begin{table}[ht]
\centering
\caption{Performance of ChatGLM3-6B-base after amplifying top 5\% Neurons.}
\label{tab:performance_amplify5per}
\resizebox{\columnwidth}{!}{
\begin{tabular}{cccc}
\toprule
Task / Metric & Normal & ICA & LLM Localization \\
\midrule
CoLA (MCC)   & 0.1497 & \textbf{0.1580} & 0.1577 \\
MNLI         & 0.5636 & \textbf{0.5638} & 0.5633 \\
MRPC (F1)    & 0.8415 & \textbf{0.8436} & \textbf{0.8436} \\
QNLI         & 0.8223 & \textbf{0.8223} & 0.8217 \\
QQP (F1)     & 0.8065 & 0.8062 & 0.8064 \\
RTE          & 0.7581 & \textbf{0.7617} & 0.7581 \\
SST-2        & 0.9553 & 0.9553 & 0.9553 \\
WNLI         & 0.6479 & 0.6338 & 0.6338 \\
\midrule
Average      & 0.6931 & 0.6931 & 0.6925 \\
\midrule
Wikitext (PPL) &10.1177 & 10.1168 & 10.1181\\
\bottomrule
\end{tabular}
}
\end{table}

\begin{figure}[ht]
    \begin{center}
    \centerline{\includegraphics[width=\columnwidth]{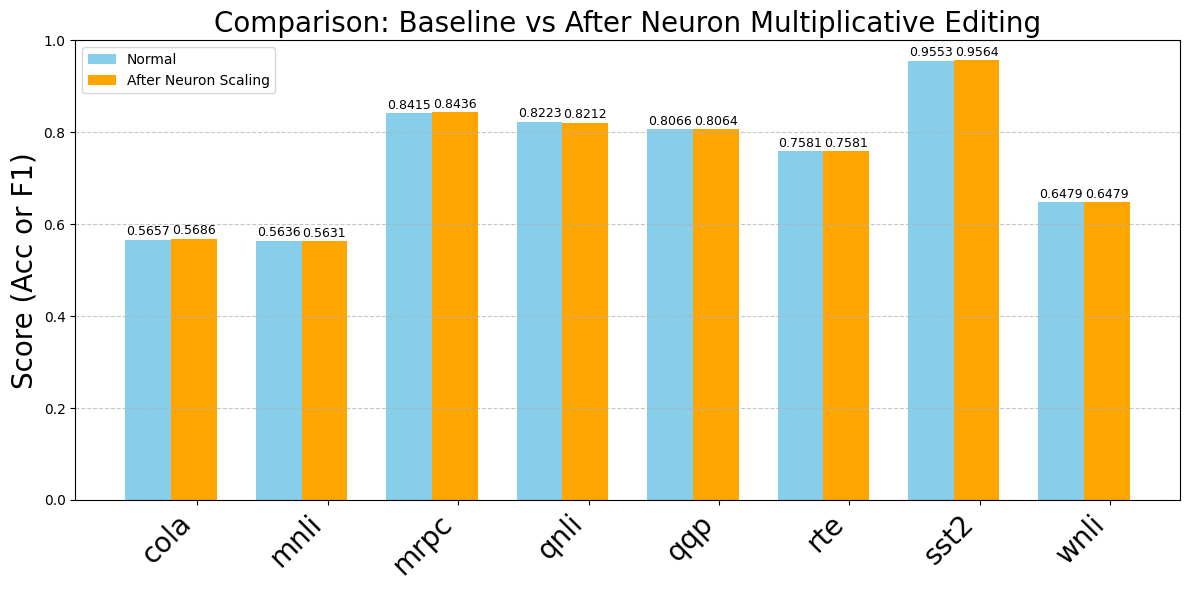}}
    \caption{Zero-shot results for ChatGLM3-6B-base after amplifying the neurons corresponding to a functional network obtained from SST-2 dataset.}
    \label{fig::glm_mul_best}
    \end{center}
\end{figure}

We first verify whether the neurons identified by our method are indeed more critical. The results of amplifying the top 5\% of functional network neurons in ChatGLM3-6B-base are shown in the Table~\ref{tab:performance_amplify5per}. Our method selects the top 5\% of neurons with the highest weights across 10 functional networks and amplifies their activations. As shown in Table~\ref{tab:performance_amplify5per}, this leads to performance improvements on several tasks, and the degree of enhancement achieved by our method exceeds that of other approaches (LLM Localization~\cite{alkhamissi2024llm}). Moreover, it can be observed that our method improves performance on the majority of tasks. Figure~\ref{fig::glm_mul_best} shows the results of applying multiplicative editing to the neurons of one functional network. The evaluation metrics used are accuracy and F1-score. After editing, the average score is 0.7457, compared to 0.7451 for the unedited (normal) model. 

Notably, editing this functional network leads to improved performance on SST-2. Among our 10 functional networks, only two enhance SST-2 performance, and these two exhibit relatively higher spatial similarity (reaching 0.1514) compared to all other pairs, whose maximum spatial similarity is only 0.0851. 

We reasonably hypothesize that the neurons shared by these two functional networks are directly involved in the sentiment analysis task on SST-2. To verify this, we took the union of the two networks and amplified their activations, which successfully improved SST-2 performance. Furthermore, even when we amplified only the non-overlapping (network-specific) neurons (those not shared between the two networks) performance on SST-2 still increased. And amplifying neurons in the other functional networks does not improve SST-2 performance. This demonstrates that the neurons in both decomposed functional networks are meaningfully associated with sentiment analysis, rather than being redundant or irrelevant.

\subsection{Do Different Methods Find the Same Important Neurons?}
We further evaluated the consistency of important neurons identified by different methods by computing the spatial similarity (IoU) between the neuron sets derived by our ICA-based approach and those from other methods. For each method, we extracted the top 5\% of neurons ranked by importance and calculated the IoU between these binary neuron masks. 

\begin{table}[ht]
\centering
\caption{Spatial similarity between the important neurons derived by our ICA-base approach and LLM Localization~\cite{alkhamissi2024llm}.}
\label{tab:iou-methods}
\resizebox{\columnwidth}{!}{
\begin{tabular}{cccc}
\toprule
Methods & ICA vs LLM Localization \\
\midrule
Spatial Similarity (IoU) & 0.0220\\
\bottomrule
\end{tabular}
}
\end{table}

Table~\ref{tab:iou-methods} shows the spatial similarity between the sets of important neurons identified by our method and those from LLM Localization. It can be observed that the two methods yield neuron sets with almost no spatial overlap, indicating that the important neurons they identify are largely distinct in location. Combined with the results in Table~\ref{tab:performance_amplify5per} and Table~\ref{tab::glm_10_5per_mask} (see in the appendix), our method identifies neurons that have a significantly greater impact on model performance than LLM Localization. This strongly supports the neuroscientific insight that functionally relevant computation in neural systems arises not from isolated neurons but from co-activated ensembles--that is, groups of neurons working together in coordinated patterns. 

\section{Discussion and Conclusion}
This study introduces a framework for analyzing LLMs by adapting analytical methods originally developed in cognitive neuroscience to identify functional brain networks from neuroimaging data. Specifically, we apply group-level ICA to neuron activations in LLMs, treating them as high-dimensional signals analogous to fMRI time series.
 
Our data-driven approach reveals reproducible ensembles of co-activating neurons within the model’s MLP layers. These ensembles form stable, task-general functional units: their collective activity persists across diverse inputs, and targeted perturbation of these groups leads to measurable degradation in representation quality. This suggests that LLMs rely on structured, coordinated functional networks for encoding semantic information, offering a system-level perspective that goes beyond isolated neuron interpretations.

Our work demonstrates that methods from cognitive neuroscience can be effectively repurposed to locate structured, task-relevant neuron ensembles in LLMs. These functional networks are not arbitrary collections of units; rather, they behave as integrated circuits, with constituent neurons jointly encoding specific semantic or computational functions. This coordinated activity persists across tasks and examples, suggesting that information processing in LLMs relies on the collective behavior of these ensembles. Recognizing this collaborative structure provides a more meaningful lens for interpreting model internals than isolated neuron analyses or heuristic importance metrics.

Although our study focuses on analysis rather than engineering applications, the clear functional role of these neuron ensembles opens a plausible avenue for future work, such as using functional networks as a guide for model compression and model fingerprint. For instance, preserving neurons within critical functional groups while simplifying or removing others could offer a more principled basis for pruning than current heuristics. The unique functional activity of neurons within functional networks, along with the functional connectivity between these networks, holds promise as a model fingerprint (similar to the conception "brain fingerprints" in neuroscience \cite{finn2015functional}) and can serve as an effective mechanism for protecting the intellectual property of large language models. Ultimately, this work underscores how cross-disciplinary frameworks, borrowing rigorously from neuroscience, can yield actionable insights into the inner mechanisms of LLMs.

\section*{Limitations}
Our investigation focused solely on the MLP layers. Future work could extend this approach to other components within these models, such as attention mechanisms or embedding layers. By broadening the scope of analysis to include these additional modules, we can gain a more comprehensive understanding of how functional networks manifest in different components of LLMs.

The algorithm used in this research is ICA, which has various derivatives. These variants can improve model's performance based on the unique characteristics of the dataset. In addition to ICA, other approaches such as dictionary learning and deep learning–based autoencoders can also be employed to decompose the internal representations of LLMs into functional networks. Consequently, there is potential for the development of novel algorithms tailored specifically to the properties of LLMs. 

\bibliography{custom}

\appendix

\section{Appendix}
\label{sec:appendix}

\subsection{ICA}
FastICA \cite{hyvarinen2000independent} is an efficient algorithm for implementing ICA and is the method used in this paper to derive FBNs from LLMs. The FastICA algorithm can be described as follows:

Pre-whitening: The signals are first centered (zero mean) and whitened to remove any linear correlations between the variables and have unit variance.

The whitened signals $\mathbf{Z}$ can be represented as:
\begin{equation}
   \mathbf{Z} = \mathbf{E}^{-1/2} \mathbf{V}^{-1} (\mathbf{X} - \mathbb{E}[\mathbf{X}]) 
\end{equation}
where $\mathbf{V}$ and $\mathbf{E}$ are the eigenvectors and eigenvalues of the covariance matrix $\mathbf{\Sigma}$ of $\mathbf{X}$.

Finding Independent Components: For each independent component $\mathbf{w}_i$, we maximize the following objective function:
\begin{equation}
    J(\mathbf{w}) = [\mathbb{E}\{G(\mathbf{w}^T \mathbf{z})\}] - \frac{1}{2} \mathbb{E}\{(\mathbf{w}^T \mathbf{z})^2\}
\end{equation}

where $G$ is a non-linear function used to approximate negentropy, $a_1$ is a constant usually $a_1 \in [1, 2]$, Common choices for $G$ include:

\begin{equation}
    G(u) = \log \cosh(a_1 u)
\end{equation}
\begin{equation}
    G(u) = -\exp(-u^2 / 2)
\end{equation}

To find the optimal $\mathbf{w}$, FastICA uses fixed-point iteration:

\begin{equation}
    \mathbf{w}_{\text{new}} = \mathbb{E}\{\mathbf{z} g(\mathbf{w}^T \mathbf{z})\} - \mathbb{E}\{g'(\mathbf{w}^T \mathbf{z})\} \mathbf{w}
\end{equation}

where $g$ is the derivative of $G$. After each iteration, normalize $\mathbf{w}$:

\begin{equation}
    \mathbf{w} \leftarrow \frac{\mathbf{w}}{\|\mathbf{w}\|}
\end{equation}

If multiple independent components need to be extracted, perform orthogonalization to ensure that the weight vectors remain orthogonal:

\begin{equation}
    \mathbf{W} \leftarrow (\mathbf{W} \mathbf{W}^T)^{-1/2} \mathbf{W}
\end{equation}

where $\mathbf{W}$ is the matrix that contains all weight vectors as columns. Repeat the iteration until the change in the weight vectors falls below a predefined threshold, indicating convergence. Once the demixing matrix $\mathbf{W}$ is obtained, the source signals $\mathbf{S}$ can be estimated from the whitened data $\mathbf{Z}$:
\begin{equation}
    \hat{\mathbf{S}} = \mathbf{W} \mathbf{Z}
\end{equation}

Finally, the mixing matrix $\mathbf{A}$, which represents the spatial patterns of the functional networks, can be obtained as the inverse of the matrix $\mathbf{W}$. Considering the whitening transformation applied to the signals, the mixing matrix can be computed as:

\begin{equation}
    \mathbf{A} = \mathbf{V} \mathbf{E}^{1/2} \mathbf{W}^{-1}
\end{equation}

In visualizations and experimental comparisons, the mixing matrix $\mathbf{A}$, representing the final derived functional networks, typically undergoes thresholding. This process involves setting a threshold to filter out lower values, ensuring that only the regions with significant activation are retained. This approach helps in focusing on the most relevant activations and improving the clarity of the results.

The CanICA (Canonical Independent Component Analysis) \cite{varoquaux2010group,varoquaux2010ica} algorithm is a spatial independent component analysis (sICA) method used for brain imaging data. The subsequent section presents a detailed mathematical description, concentrating on PCA preprocessing and the application of FastICA for spatial component extraction.

Assume we have an observation data matrix $X \in \mathbb{R}^{n \times t}$, where $n$ is the number of sensors or voxels (spatial dimension), and $t$ is the number of time points. Each column represents the observation at one time point, and each row represents the observations of one sensor or voxel across all time points.

In CanICA, Principal Component Analysis (PCA) is first applied to reduce the dimensionality of the data. PCA can be implemented via Singular Value Decomposition (SVD):
$$ 
X = U \Sigma V^T
$$
where:
\begin{itemize}
    \item $U \in \mathbb{R}^{n \times n}$ is the left singular vector matrix.
    \item  $\Sigma \in \mathbb{R}^{n \times t}$ is the singular value matrix, with singular values on its diagonal.
    \item  $V \in \mathbb{R}^{t \times t}$ is the right singular vector matrix.
\end{itemize}
 
We select the top $k$ singular vectors corresponding to the largest singular values to construct the reduced data matrix $X_{\text{reduced}}$:
$$
X_{\text{reduced}} = U_k \Sigma_k
$$
where $U_k$ and $\Sigma_k$ are the first $k$ columns of $U$ and the first $k$ rows of $\Sigma$, respectively.

FastICA is then applied to the reduced data. Unlike traditional ICA, the mixing matrix $A$ here does not represent the functional networks. In spatial ICA $S \in \mathbb{R}^{k \times t}$, which is the source signal matrix that represents the functional networks that are to be extracted. The model can be expressed as:
$$ 
X_{\text{reduced}} = A S 
$$
where $A \in \mathbb{R}^{n \times k}$ is the mixing matrix, indicating how spatial components combine to form the observed data.

In practice, the reduced data $X_{\text{reduced}}$ is fed into the FastICA algorithm for further processing. 

The extracted spatial components are stored in the source signal matrix $S$, with each row vector representing an independent spatial pattern.

\subsection{Stable Functional Networks within LLM}
\label{appendix::stable-FBN}
The experiments in Section~\ref{sec::consistent} are shown in Table~\ref{tab::qwen_cnt}, ~\ref{tab::qwen7b_cnt} and ~\ref{tab::glm_cnt}.

\begin{table}[t]
\caption{The number of the counterparts in 100 individual-level analysis for the functional network templates in group-level analysis in Qwen2.5-3B-Instruct.}
\label{tab::qwen_cnt}
\begin{center}
\begin{small}
\begin{sc}
\begin{tabular}{ccccc}
\toprule
Templates & News & Wiki & Math & code \\
\midrule
1 & 0  & 57 & 0  & 4  \\
2 & 7  & 51 & 0  & 62 \\
3 & 0  & 44 & 5  & 80 \\
4 & 11 & 53 & 6  & 21 \\
5 & 51 & 17 & 0  & 12 \\
6 & 19 & 20 & 0  & 26 \\
7 & 74 & 89 & 48 & 62 \\
8 & 8  & 10 & 9  & 61 \\
9 & 14 & 0  & 0  & 54 \\
10 & 56 & 88 & 7  & 60 \\
\bottomrule
\end{tabular}
\end{sc}
\end{small}
\end{center}
\end{table}

\begin{table}[t]
\caption{The number of the counterparts in 100 individual-level analysis for the functional network templates in group-level analysis in Qwen2.5-7B-Instruct. }
\label{tab::qwen7b_cnt}
\begin{center}
\begin{small}
\begin{sc}
\begin{tabular}{ccccc}
\toprule
Templates & News & Wiki & Math & code \\
\midrule
1 & 12  & 71  & 121 & 60 \\
2 & 4   & 10  & 115 & 60 \\
3 & 89  & 25  & 139 & 64 \\
4 & 0   & 2   & 0   & 131 \\
5 & 19  & 2   & 139 & 73 \\
6 & 67  & 1   & 0   & 61 \\
7 & 131 & 62  & 22  & 75 \\
8 & 23  & 7   & 4   & 5 \\
9 & 31  & 87  & 7   & 68 \\
10 & 74 & 19  & 191 & 119 \\
\bottomrule
\end{tabular}
\end{sc}
\end{small}
\end{center}
\end{table}

\begin{table}[t]
\caption{The number of the counterparts in 100 individual-level analysis for the functional network templates in group-level analysis in ChatGLM3-6B-base.}
\label{tab::glm_cnt}
\begin{center}
\begin{small}
\begin{sc}
\begin{tabular}{ccccc}
\toprule
Templates & News & Wiki & Math & code \\
\midrule
1  & 0  & 47 & 0  & 50 \\
2  & 27 & 5  & 8  & 2  \\
3  & 4  & 10 & 0  & 0  \\
4  & 8  & 88 & 0  & 38 \\
5  & 0  & 33 & 6  & 61 \\
6  & 27 & 75 & 13 & 202 \\
7  & 3  & 0  & 12 & 11 \\
8  & 26 & 0  & 1  & 0  \\
9  & 12 & 16 & 0  & 27 \\
10 & 0  & 51 & 0  & 171 \\
\bottomrule
\end{tabular}
\end{sc}
\end{small}
\end{center}
\end{table}

\subsection{Distribution of Neurons in Functional Networks}
\begin{figure*}[ht]
\begin{center}
\centerline{\includegraphics[width=\textwidth]{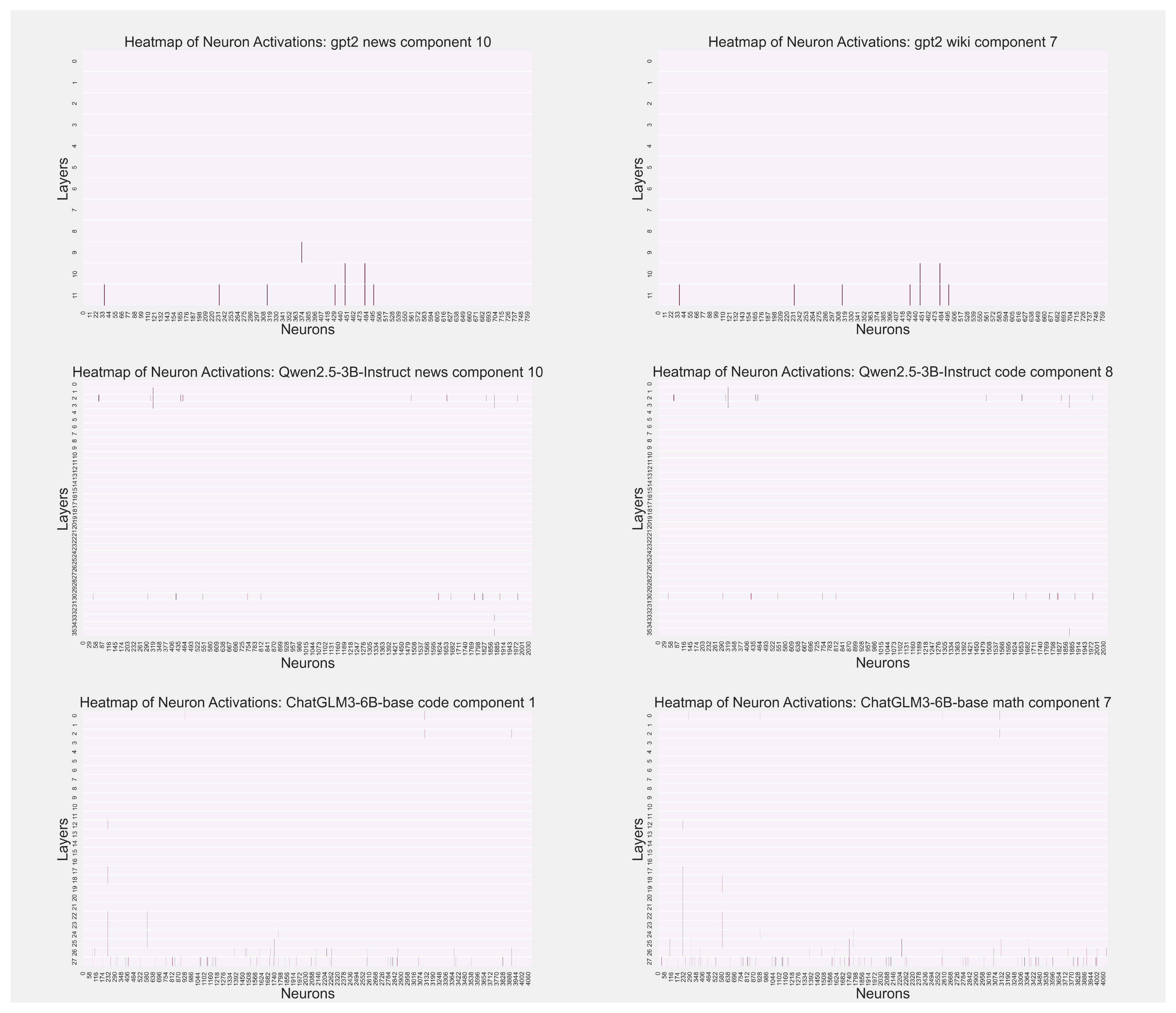}}
\caption{Example functional networks that are consistent across various types of input samples in group level analysis. Each row represents the neurons in an MLP layer where activated neurons are highlighted.}
\label{fig::fbn}
\end{center}
\end{figure*}

In Figure~\ref{fig::fbn}, we present several group-level functional networks derived from different datasets. These functional networks involve only a small fraction of total neurons, ranging from less than 0.1\% to around 2\%, consistent with the well-established principle in neuroscience that functional brain networks also activate a sparse subset of regions \cite{dong2020discovering,liu2023spatial} (see Fig.~\ref{fig::frame}(a) for an illustration of human functional networks). This sparsity aligns with our observation that LLMs, like the human brain, rely on highly selective neural activation. Moreover, Figure~\ref{fig::fbn} visually demonstrates that similar functional network structures emerge across diverse input data types, supporting our claim that stable, reusable functional units exist in LLMs.

\subsection{Functional Networks Editing Experiments}
The neuron-leison experiments results are shown in Table~\ref{tab::qwen7b_10_all} and ~\ref{tab::glm_10_all}. Table~\ref{tab::glm_10_5per_mask} shows the results of experiments on ChatGLM3-6B-base using the lm-eval framework, comparing our method (where the top 5\% most important neurons are inhibited) with the LLM Localization~\cite{alkhamissi2024llm} approach. It can be observed that, when inhibiting the same number of neurons, our method causes a greater drop in model performance, indicating that the neurons identified by our approach are more critical to the model’s functionality.

\begin{table}[ht]
\caption{Performance of ChatGLM3-6B-base after inhibiting top 5\% Neurons.}
\label{tab::glm_10_5per_mask}
\begin{center}
\resizebox{\columnwidth}{!}{
\begin{small}
\begin{sc}
\begin{tabular}{cccc}
\toprule
Task & Normal & ICA & LLM Localization \\
\midrule
CoLA (MCC)   & 0.1497 & -0.0179 & 0.0702 \\
MNLI         & 0.5636 & 0.3533  & 0.4034 \\
MRPC (F1)    & 0.8415 & 0.0676  & 0.7470 \\
QNLI         & 0.8223 & 0.5318  & 0.5535 \\
QQP (F1)     & 0.8065 & 0.5334  & 0.4244 \\
RTE          & 0.7581 & 0.5632  & 0.6570 \\
SST-2        & 0.9553 & 0.5952  & 0.7787 \\
WNLI         & 0.6479 & 0.4648  & 0.4225 \\
\midrule
Average      & 0.6931 & 0.3864  & 0.5071 \\
\midrule
Wikitext (PPL) &10.1177 & 40.4245 & 14.6199\\
\bottomrule
\end{tabular}
\end{sc}
\end{small}
}
\end{center}
\end{table}

\begin{figure}[ht]
\begin{center}
\centerline{\includegraphics[width=\columnwidth]{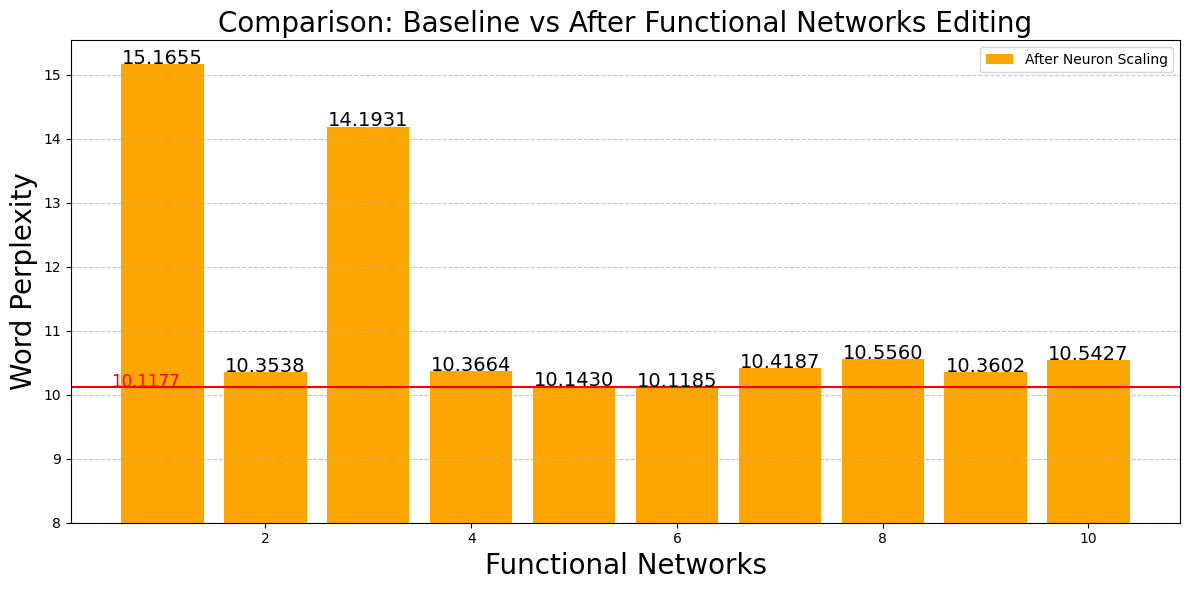}}
\caption{Results perplexity of additive editing for 10 functional networks in ChatGLM3-6B-base. Adding value is set as 1.}
\label{fig::add_glm}
\end{center}
\end{figure}

\begin{figure}[ht]
\begin{center}
\centerline{\includegraphics[width=\columnwidth]{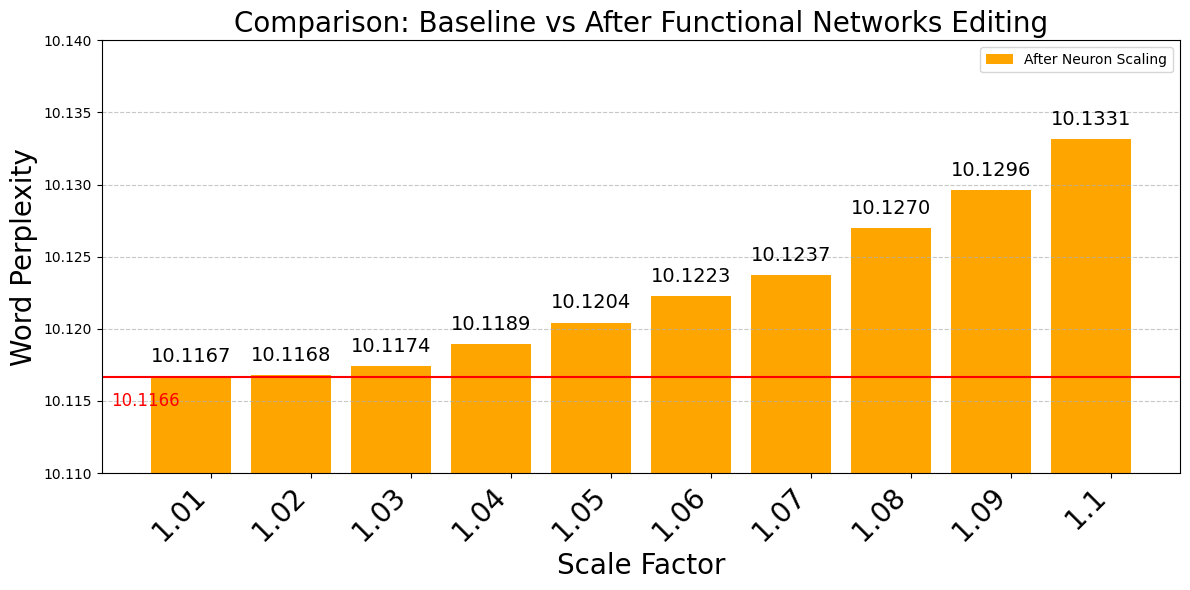}}
\caption{Results perplexity of different multiplicative factors editing in ChatGLM3-6B-base.}
\label{fig::mul_glm}
\end{center}
\end{figure}

\begin{figure}[ht]
\begin{center}
\centerline{\includegraphics[width=\columnwidth]{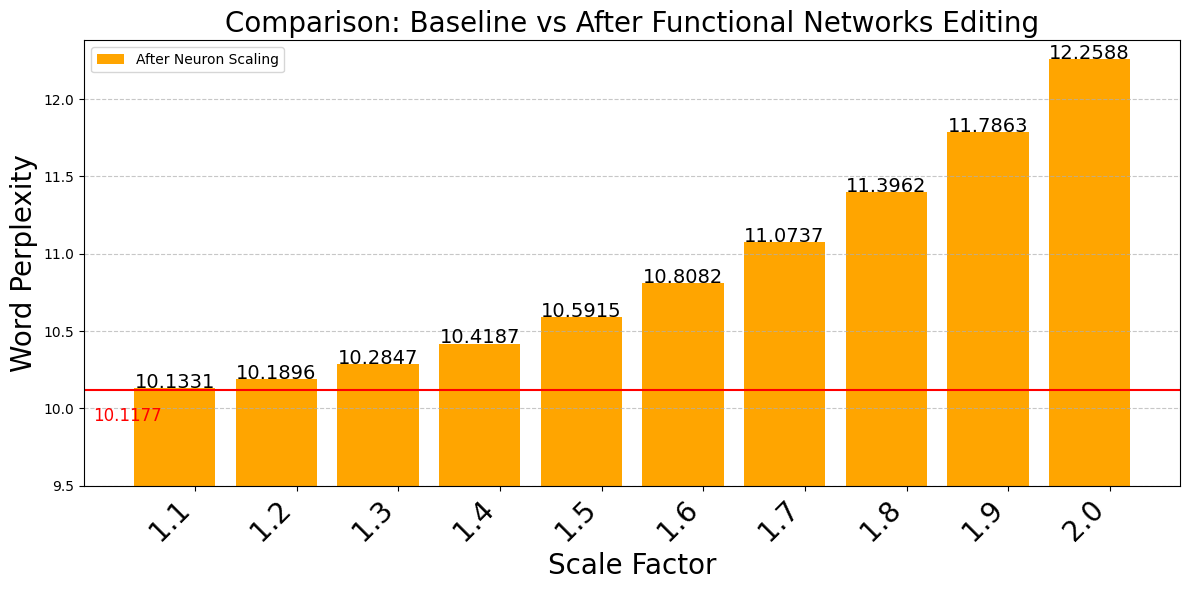}}
\caption{Results perplexity of different multiplicative factors editing in ChatGLM3-6B-base.}
\label{fig::mul_glm_1.1}
\end{center}
\end{figure}

The results of additive editing and other multiplicative factors in ChatGLM3-6B-base are shown in Figure~\ref{fig::add_glm}, ~\ref{fig::mul_glm} and ~\ref{fig::mul_glm_1.1}. If values greater than 1 (e.g., between 1.5 and 3) are added to the neurons of functional networks in ChatGLM3-6B-base, the model’s word perplexity can surge to tens of thousands or higher. It can be observed that neither additive nor multiplicative editing should use excessively large values, as this severely degrades model performance. Moreover, multiplicative editing yields more stable results, making it the preferred approach for neuron editing. 

\begin{table}[t]
\caption{Performance of Qwen2.5-7B-Instruct after inhibiting functional networks. First Column: Dataset. Second Column: Model performance under normal conditions (without inhibiting). Third Column: Model performance after inhibiting 15\% of the neurons randomly. Fourth Column: Model performance after inhibiting the neurons belonging to 10 specific functional networks.}
\label{tab::qwen7b_10_all}
\begin{center}
\resizebox{\columnwidth}{!}{
\begin{small}
\begin{sc}
\begin{tabular}{cccc}
\toprule
Datasets & Normal &  Masked 15\% & Masked 10\\
& & (15053) & (361-2217)\\
\midrule
COLA       & 0.7641 & 0.7143 & 0.0000 \\
MRPC       & 0.6765 & 0.3873 & 0.1691 \\
SST-2      & 0.9358 & 0.9094 & 0.0000 \\
MNLI       & 0.7243 & 0.6052 & 0.0000 \\
QNLI       & 0.8060 & 0.5629 & 0.0000 \\
QQP        & 0.8388 & 0.7367 & 0.0000 \\
RTE        & 0.8448 & 0.8339 & 0.0000 \\
WNLI       & 0.8169 & 0.7746 & 0.0000 \\
\bottomrule
\end{tabular}
\end{sc}
\end{small}
}
\end{center}
\end{table}

\begin{table}[t]
\caption{Performance of ChatGLM3-6B-base after inhibiting Functional Networks.}
\label{tab::glm_10_all}
\begin{center}
\resizebox{\columnwidth}{!}{
\begin{small}
\begin{sc}
\begin{tabular}{cccc}
\toprule 
Datasets & Normal & Masked 15\% & Masked 10 \\
         &        & (17203)     &           \\
\midrule
COLA       & 0.6893  & 0.6913 & 0.0000 \\
MRPC       & 0.7843  & 0.8015 & 0.0000 \\
SST-2      & 0.9553  & 0.9346 & 0.0000 \\
MNLI       & 0.5634  & 0.4832 & 0.0000 \\
QNLI       & 0.8223  & 0.7633 & 0.0000 \\
QQP        & 0.8548  & 0.8392 & 0.6335 \\
WNLI       & 0.6479  & 0.5352 & 0.5493 \\
RTE        & 0.7581  & 0.7040 & 0.4188 \\
AG NEWS    & 0.9128  & 0.8961 & 0.0000 \\
SQUAD      & 0.9021  & 0.8864 & 0.0000 \\
\bottomrule 
\end{tabular}
\end{sc}
\end{small}
}
\end{center}
\end{table}

\begin{figure}[ht]
\begin{center}
\centerline{\includegraphics[width=\columnwidth]{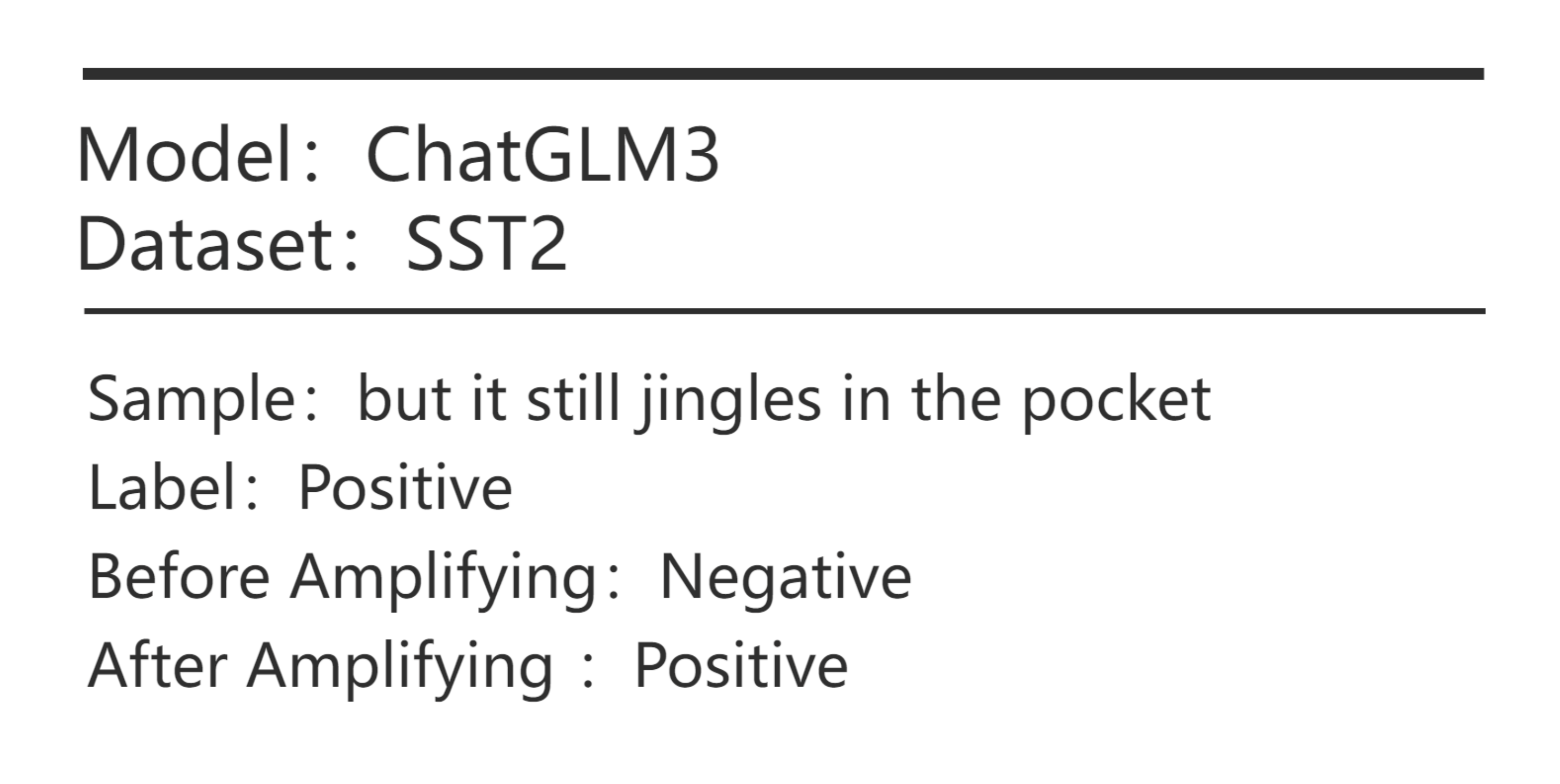}}
\caption{Correcting an SST-2 prediction by amplifying a functional network.}
\label{fig::amplify_sst2_example}
\end{center}
\end{figure}

We present an example where amplifying the neuronal signals of a functional network improves the LLM’s performance on the SST-2 dataset. As shown in Figure~\ref{fig::amplify_sst2_example}, this intervention corrects a prediction that would otherwise be incorrect under normal (unedited) model behavior—turning an originally wrong output into the correct one.

\subsection{Neuron Probing Experiment}
We investigate whether inhibiting functional networks impairs the LLM’s feature representation capability, not just its text generation. Specifically, after inhibiting certain functional networks, we observe degraded generation quality (e.g., incorrect answers or gibberish). However, this could stem from disruption in the output head or decoding process rather than damaged internal representations.

To isolate the effect on representation, we conduct a linear probing experiment: we freeze the LLM (with inhibiting functional networks), extract features from its last layer, and train a simple linear classifier (e.g., logistic regression) on these features for a downstream task. If classification performance drops significantly compared to the normal model, it indicates that the semantic content of the representations has been compromised, not merely the generation pipeline. This provides evidence that the inhibiting functional networks are essential for maintaining meaningful feature encoding.

\begin{table}[t]
\caption{Classification Performance of ChatGLM3-6B-base Using Logistic Regression on the Last Layer Features}
\label{tab:classification_performance}
\begin{center}
\begin{small}
\begin{sc}
\begin{tabular}{cccc}
\toprule
Datasets & Normal & Masked 15\% & Masked 10 \\
         &        & (17203)   & (1827-3354) \\
\midrule
QNLI     & 0.885  & 0.88      & 0.63       \\
COLA     & 0.8038 & 0.7895    & 0.6938     \\
MRPC     & 0.8049 & 0.7683    & 0.6707     \\
MNLI     & 0.86   & 0.865     & 0.705      \\
QQP      & 0.825  & 0.85      & 0.6825     \\
SST-2    & 0.9314 & 0.9314    & 0.8514     \\
\bottomrule
\end{tabular}
\end{sc}
\end{small}
\end{center}
\end{table}

\begin{table}[t]
\caption{Classification Performance of Qwen2.5-7B Using Logistic Regression on the Last Layer Features}
\label{tab:classification_performance_qwen2.5}
\begin{center}
\begin{small}
\begin{sc}
\begin{tabular}{cccc}
\toprule
Datasets & Normal & Masked 15\% & Masked 10 \\
         &        & (15052)   & (957-2035) \\
\midrule
QNLI     & 0.9    & 0.87      & 0.635     \\
COLA     & 0.8038 & 0.7464    & 0.6746    \\
MRPC     & 0.7805 & 0.8171    & 0.6463    \\
MNLI     & 0.87   & 0.84      & 0.4425    \\
QQP      & 0.83   & 0.8225    & 0.6975    \\
SST-2    & 0.9257 & 0.9486    & 0.7486    \\
\bottomrule 
\end{tabular}
\end{sc}
\end{small}
\end{center}
\end{table}

To assess whether inhibiting functional networks impairs the feature representation capability of LLMs—distinct from mere disruption of text generation—we conduct linear probing experiments on the last-layer activations. As shown in Tables~\ref{tab:classification_performance} and~\ref{tab:classification_performance_qwen2.5}, we evaluate two models—ChatGLM3-6B-base and Qwen2.5-7B—under three conditions: (1) Normal (no inhibiting), (2) Masked 15\% (randomly inhibiting 15\% of neurons), and (3) Masked 10 (inhibiting 10 identified functional networks, corresponding to neuron sets ranging from ~957 to 3,354 units).

The results reveal a clear pattern. Random inhibiting of 15\% of neurons has minimal impact on downstream classification performance across all datasets, just like the results in neuron-leision experiments. In many cases, such as SST-2 for ChatGLM3-6B-base or MRPC for Qwen2.5-7B, performance even slightly improves, likely due to incidental removal of noisy or redundant units. This suggests that general neuron loss does not significantly degrade representational quality.

In stark contrast, inhibiting just 10 specific functional networks leads to substantial performance drops, confirming that these networks carry essential semantic information. For example, on the MNLI dataset, Qwen2.5-7B’s accuracy plummets from 0.87 (Normal) to 0.4425 (inhibiting 10), a 49\% relative decline, while ChatGLM3-6B-base drops from 0.86 to 0.705. Similar trends appear across QNLI, MRPC, and QQP, with classification accuracy under Masked 10 consistently falling to the 0.63–0.70 range, despite remaining above chance. This demonstrates that while the model’s feature extractor is not completely disabled, its ability to encode task-relevant semantics is significantly compromised.

Although, Qwen2.5-7B slightly stronger than ChatGLM3-6B-base in the Normal condition, exhibits greater sensitivity to functional network inhibiting, especially on complex tasks like MNLI and COLA. This implies that Qwen2.5-7B may depend more heavily on these specific functional networks for high-level reasoning, possibly reflecting architectural or training differences between the two model families.

Crucially, these probing results contrast sharply with the models’ behavior in zero-shot generation, where inhibiting the same networks often leads to complete failure (e.g., incoherent outputs). The fact that linear probes still extract usable features while autoregressive generation fails. This suggests a dual impact: (1) functional networks are vital for constructing high-quality internal representations, and (2) their disruption disproportionately affects the generative decoding process. 

These findings provide strong evidence that the identified functional networks are not incidental but core computational units that shape both representation and generation in LLMs.

\subsection{Functional Interpretation of ICA Components}
ICA decomposes the model’s internal activations into a set of statistically independent components, each representing a distinct functional network. By design, each component is expected to capture a coherent and interpretable aspect of model computation, ideally corresponding to a specific linguistic or cognitive function.

To interpret these components, we feed an input sentence through the model, apply ICA to its hidden activations, and extract the resulting source signals \( \mathbf{S} \). For a given component, the entries in its source signal are aligned with the input tokens and used as attribution weights. We then visualize these weights to highlight which tokens most strongly activate that functional network, offering insight into the component’s potential role (e.g., attending to sentiment words, entities, or syntactic structures).

\begin{figure}[ht]
\begin{center}
\centerline{\includegraphics[width=\columnwidth]{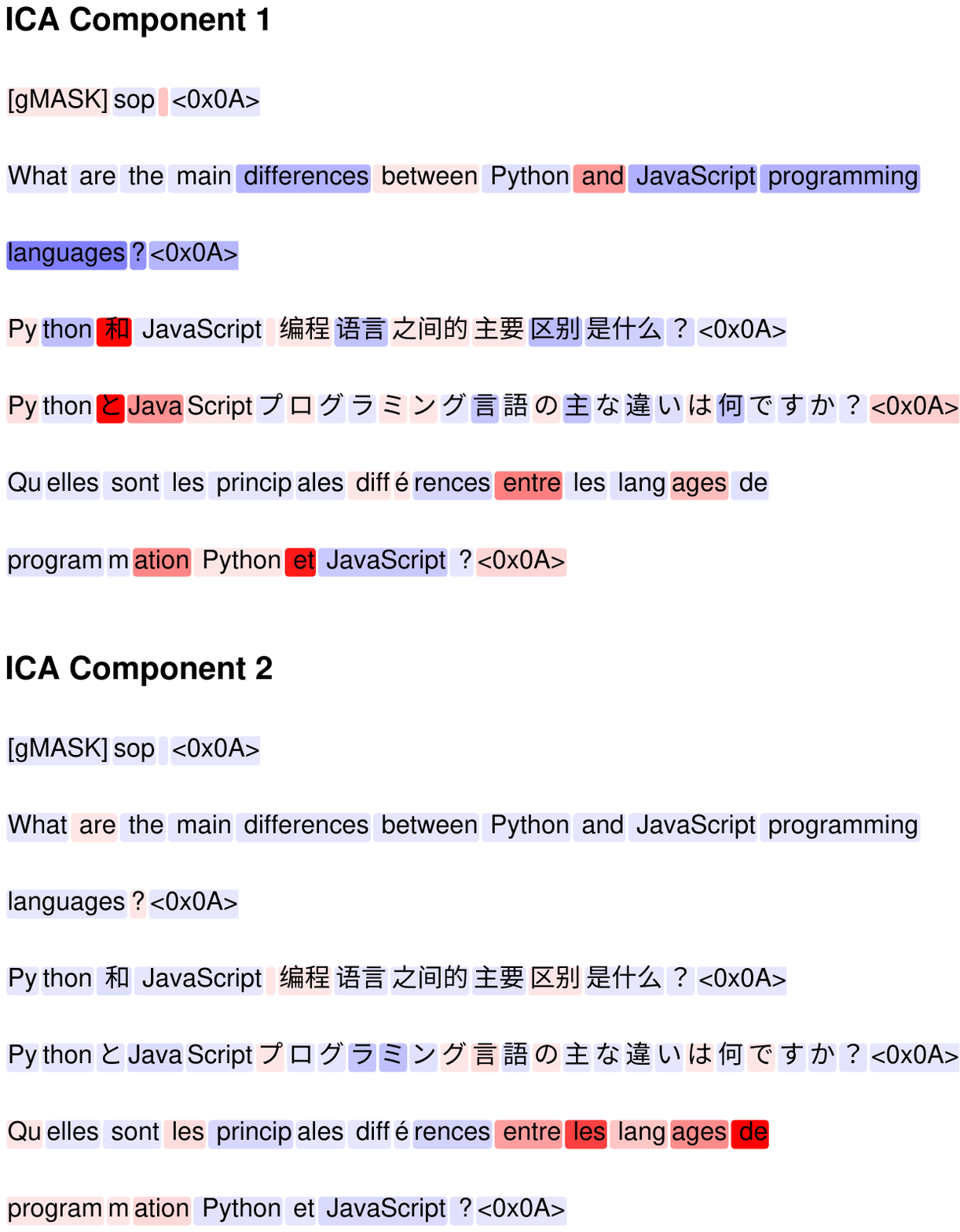}}
\caption{Interpreting ICA components by projecting their source signals back onto input tokens as attribution weights. High-weight tokens are visualized to reveal what linguistic features each functional network responds to.}
\label{fig::fbn_interpret_1}
\end{center}
\end{figure}

\begin{figure}[ht]
\begin{center}
\centerline{\includegraphics[width=\columnwidth]{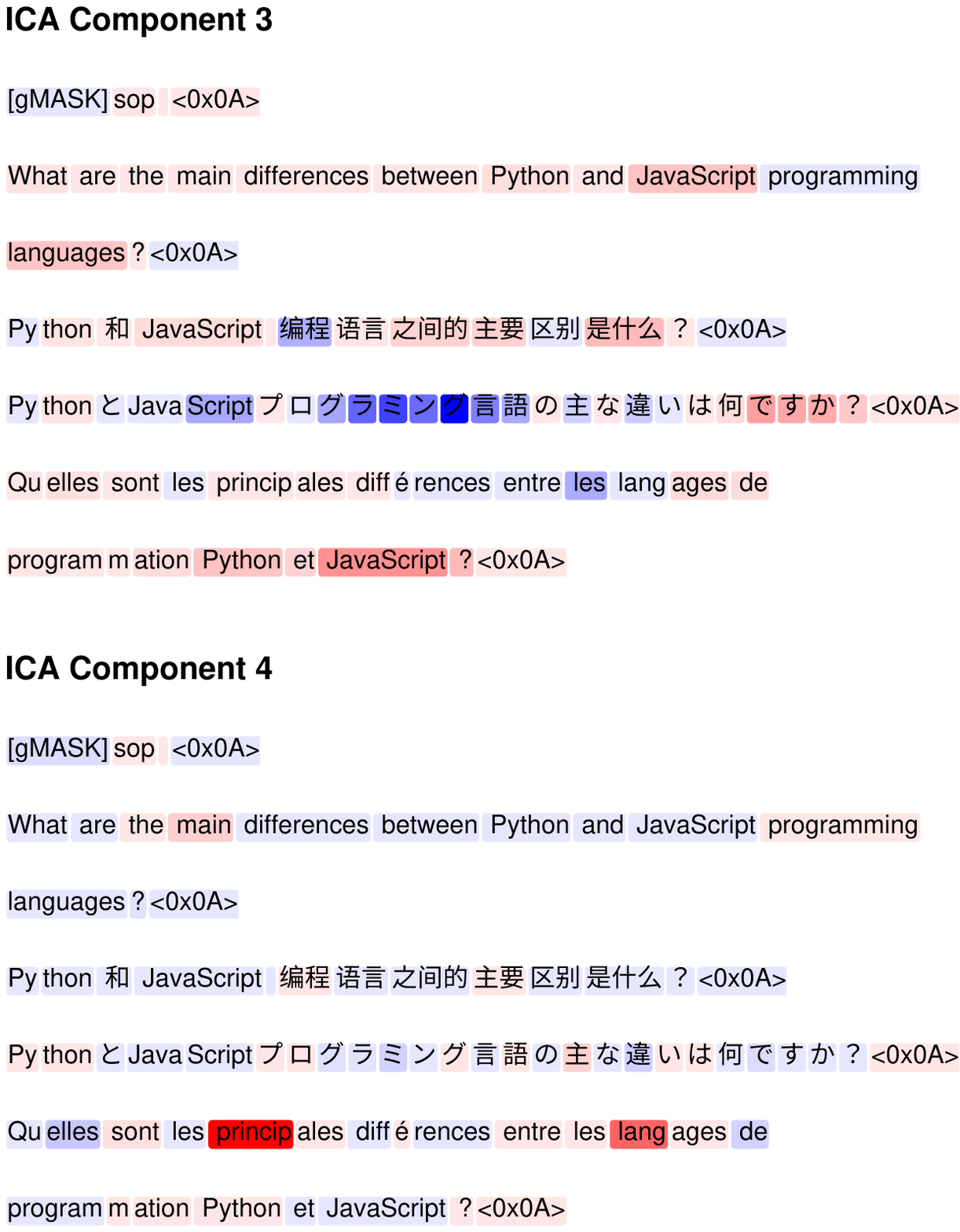}}
\caption{Interpreting ICA components by projecting their source signals back onto input tokens as attribution weights. High-weight tokens are visualized to reveal what linguistic features each functional network responds to.}
\label{fig::fbn_interpret_2}
\end{center}
\end{figure}

\begin{figure}[ht]
\begin{center}
\centerline{\includegraphics[width=\columnwidth]{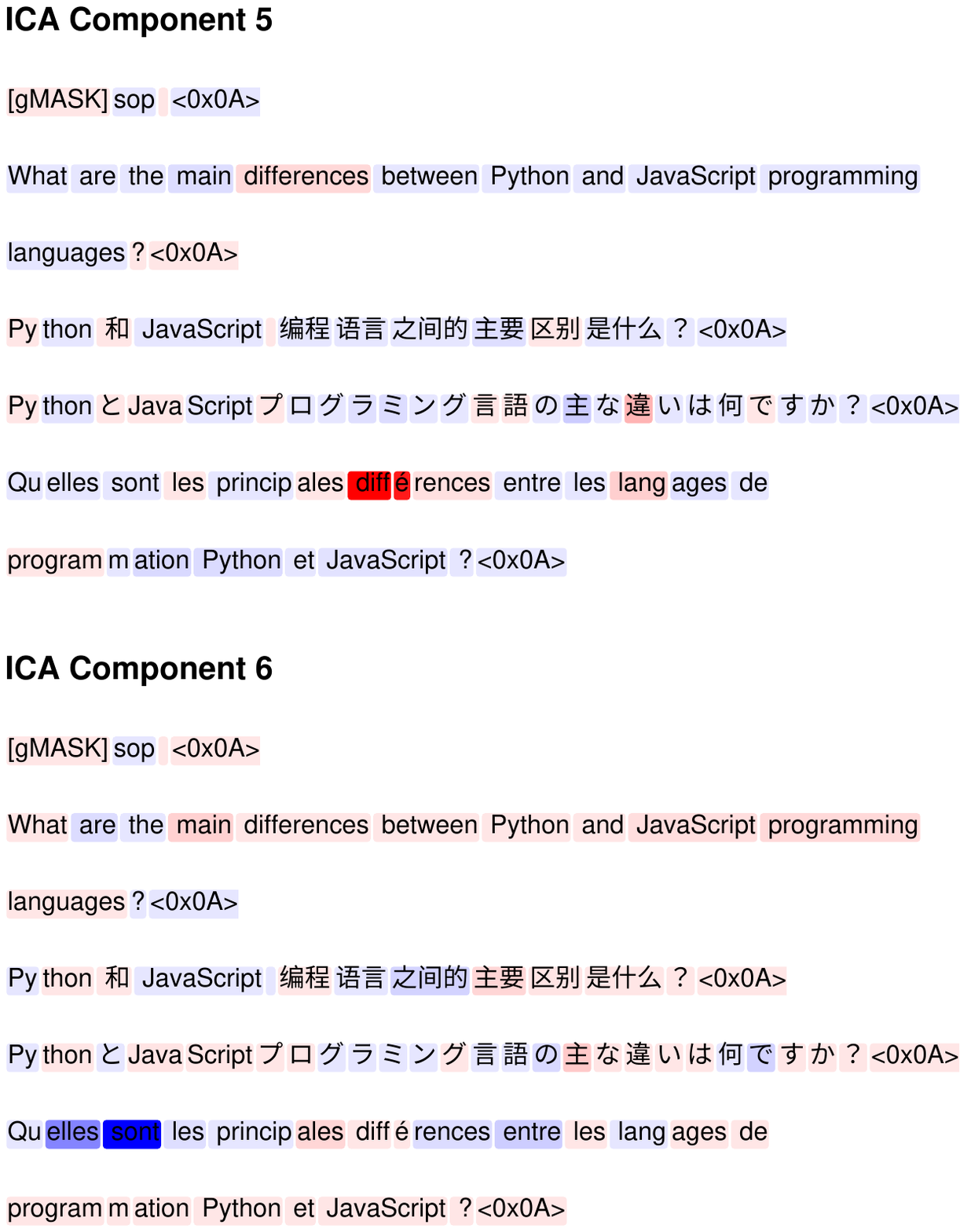}}
\caption{Interpreting ICA components by projecting their source signals back onto input tokens as attribution weights. High-weight tokens are visualized to reveal what linguistic features each functional network responds to.}
\label{fig::fbn_interpret_3}
\end{center}
\end{figure}

\begin{figure}[ht]
\begin{center}
\centerline{\includegraphics[width=\columnwidth]{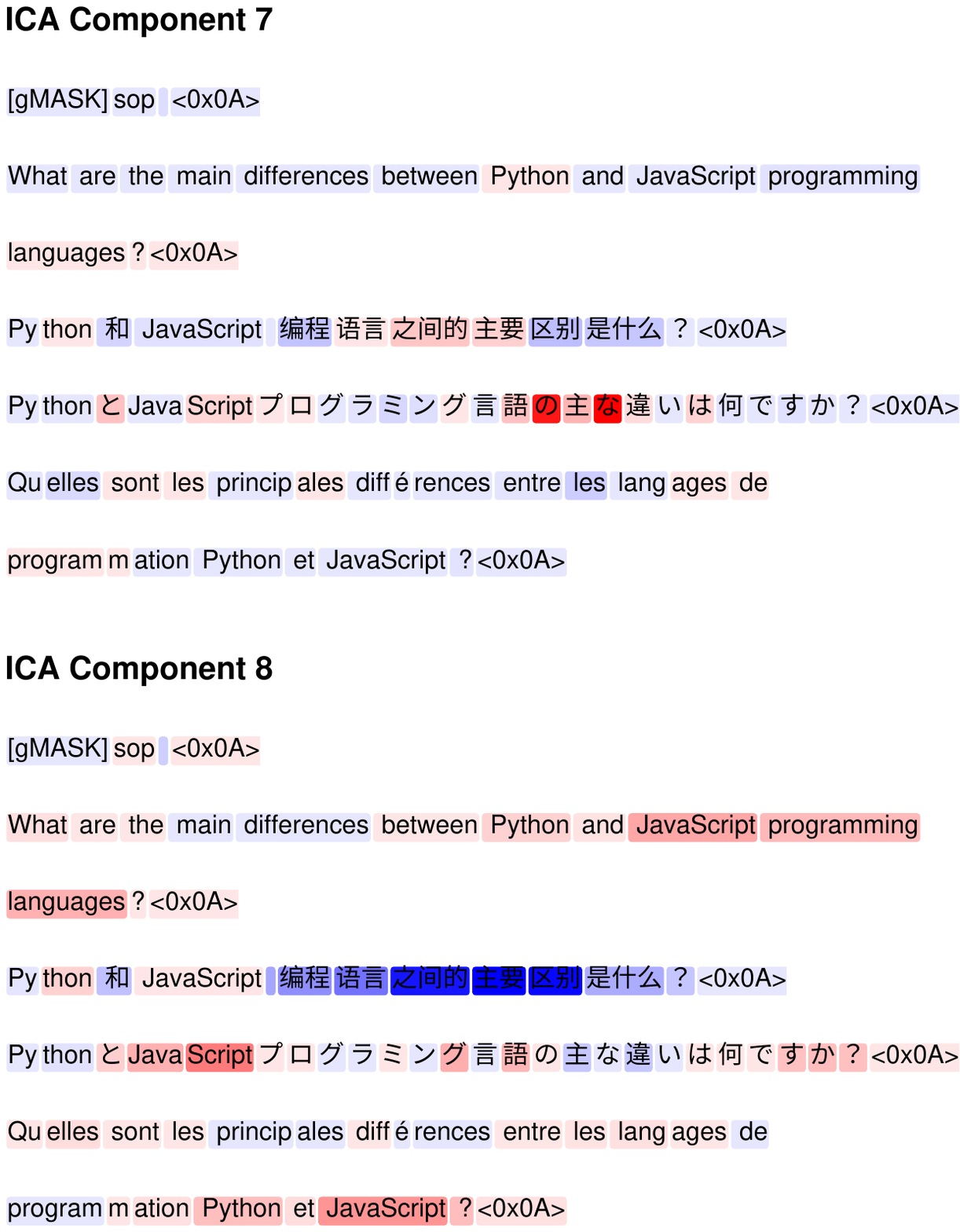}}
\caption{Interpreting ICA components by projecting their source signals back onto input tokens as attribution weights. High-weight tokens are visualized to reveal what linguistic features each functional network responds to.}
\label{fig::fbn_interpret_4}
\end{center}
\end{figure}

\begin{figure}[ht]
\begin{center}
\centerline{\includegraphics[width=\columnwidth]{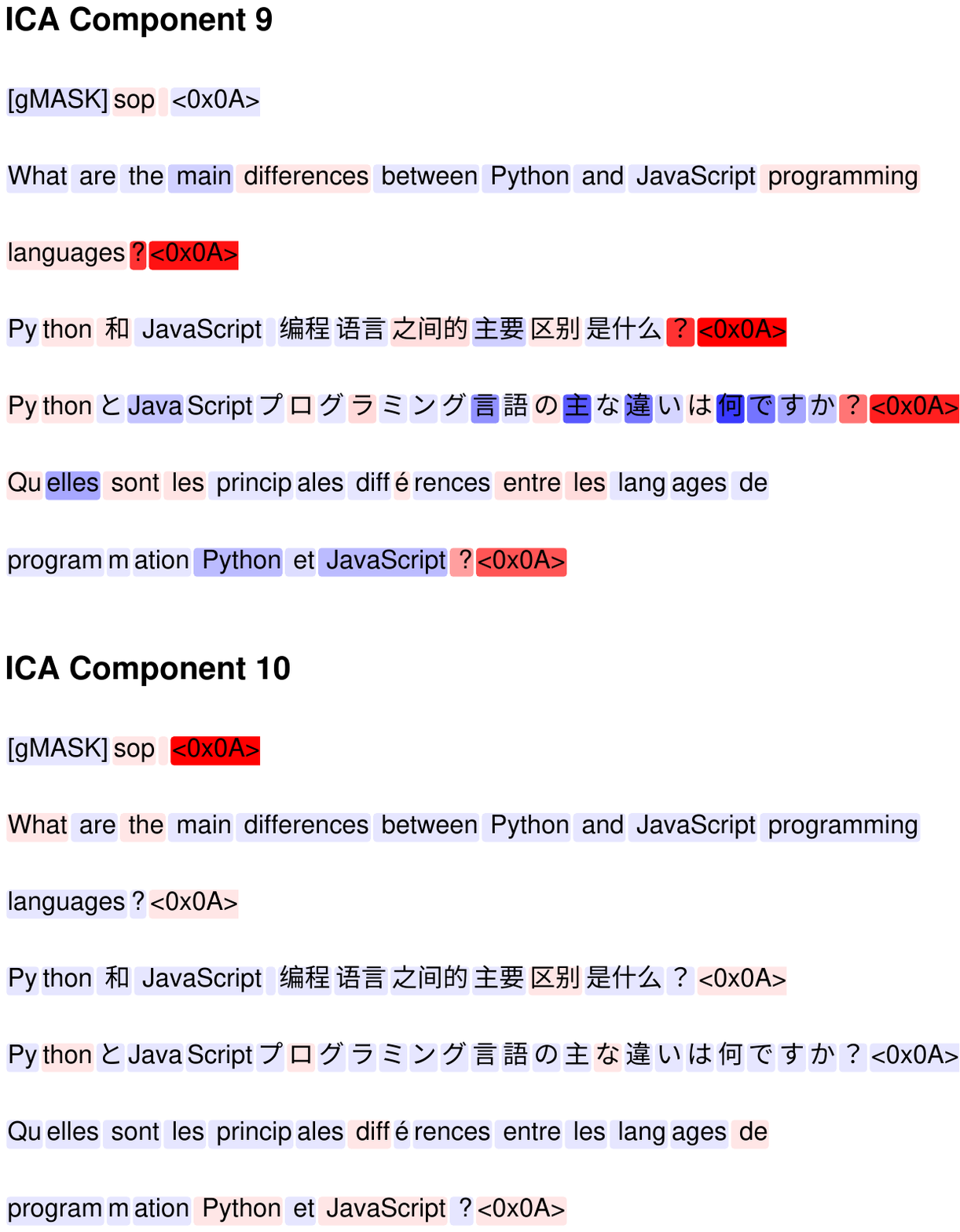}}
\caption{Interpreting ICA components by projecting their source signals back onto input tokens as attribution weights. High-weight tokens are visualized to reveal what linguistic features each functional network responds to.}
\label{fig::fbn_interpret_5}
\end{center}
\end{figure}

\begin{figure}[ht]
\begin{center}
\centerline{\includegraphics[width=\columnwidth]{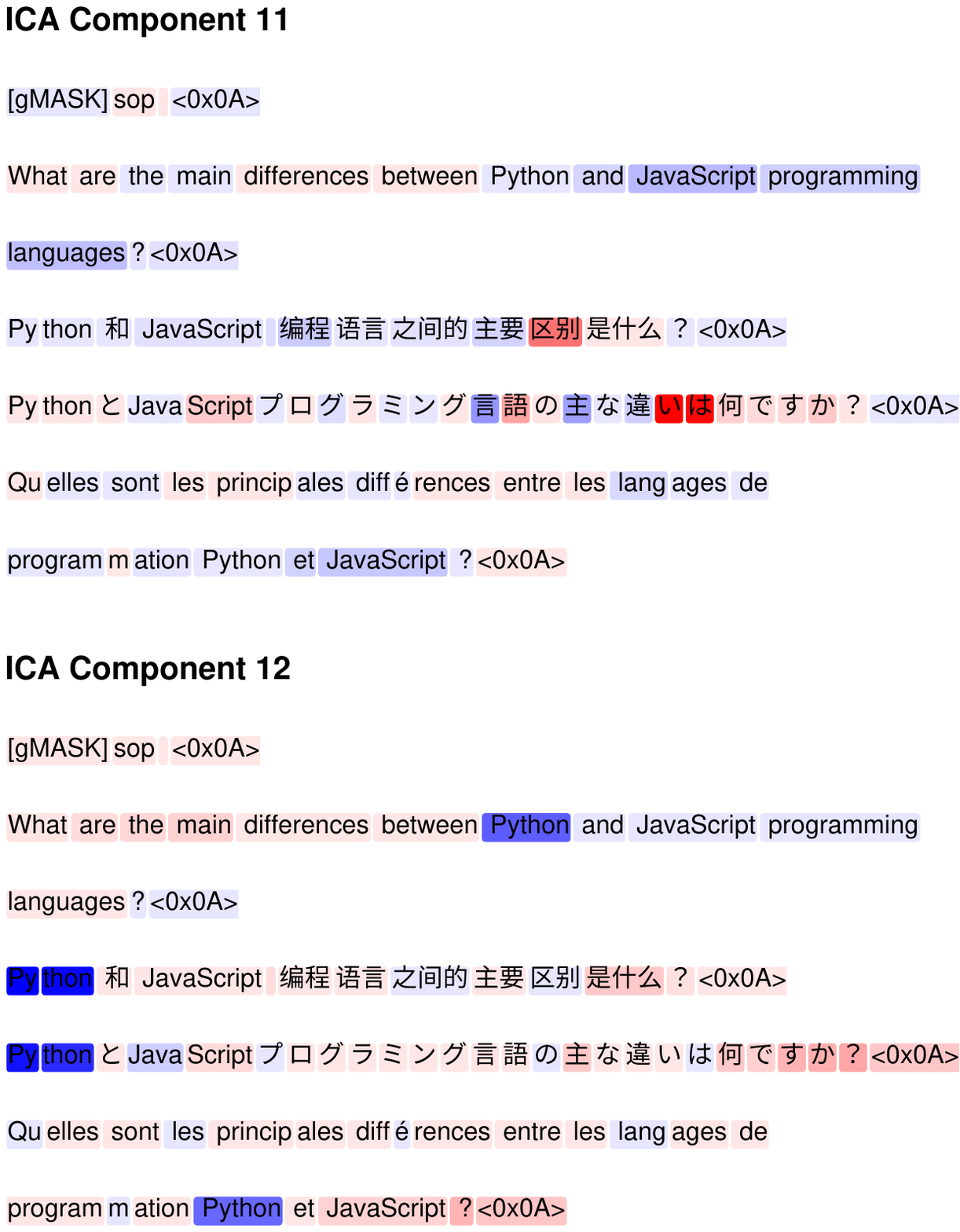}}
\caption{Interpreting ICA components by projecting their source signals back onto input tokens as attribution weights. High-weight tokens are visualized to reveal what linguistic features each functional network responds to.}
\label{fig::fbn_interpret_6}
\end{center}
\end{figure}

\begin{figure}[ht]
\begin{center}
\centerline{\includegraphics[width=\columnwidth]{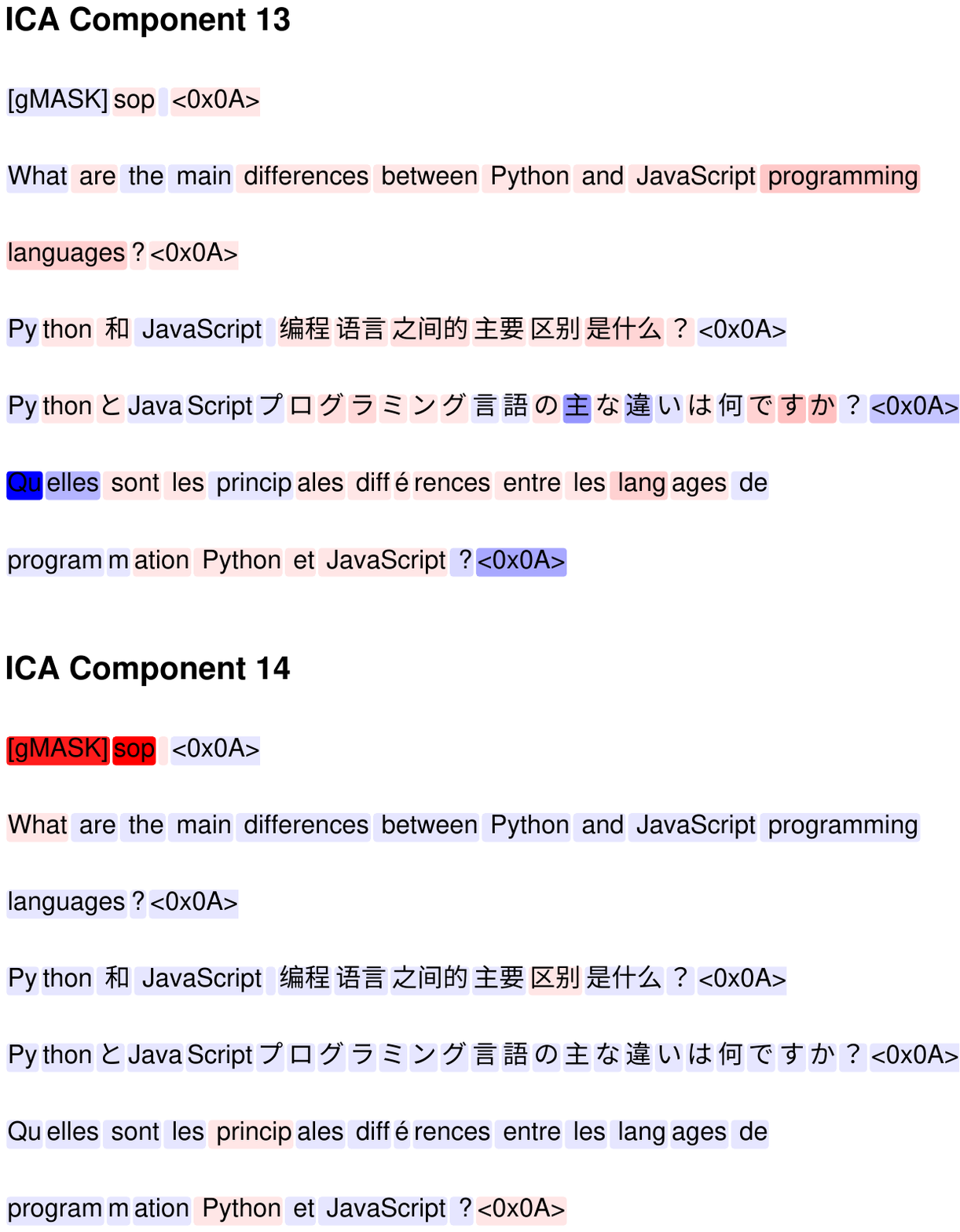}}
\caption{Interpreting ICA components by projecting their source signals back onto input tokens as attribution weights. High-weight tokens are visualized to reveal what linguistic features each functional network responds to.}
\label{fig::fbn_interpret_7}
\end{center}
\end{figure}

\begin{figure}[ht]
\begin{center}
\centerline{\includegraphics[width=\columnwidth]{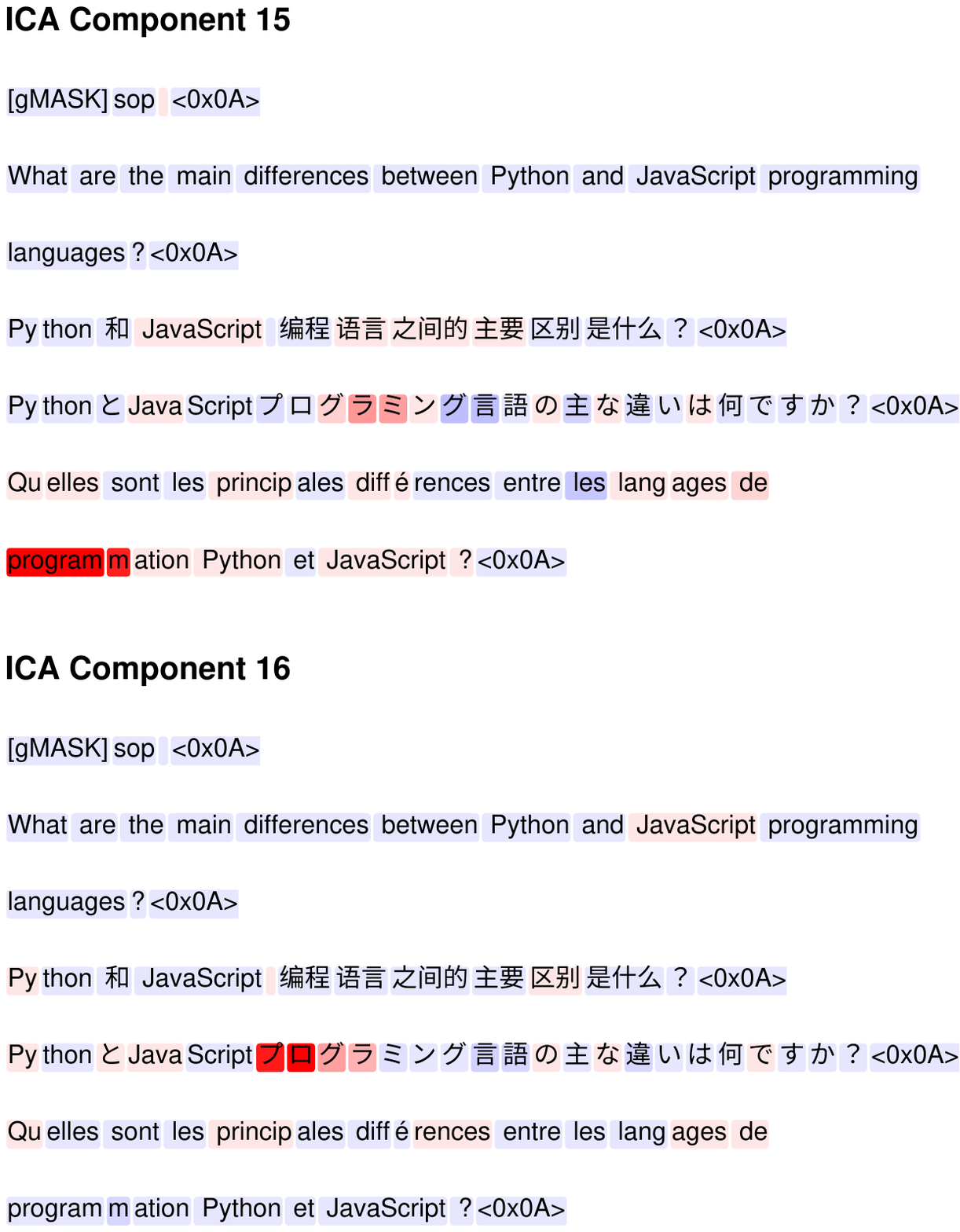}}
\caption{Interpreting ICA components by projecting their source signals back onto input tokens as attribution weights. High-weight tokens are visualized to reveal what linguistic features each functional network responds to.}
\label{fig::fbn_interpret_8}
\end{center}
\end{figure}

\begin{figure}[ht]
\begin{center}
\centerline{\includegraphics[width=\columnwidth]{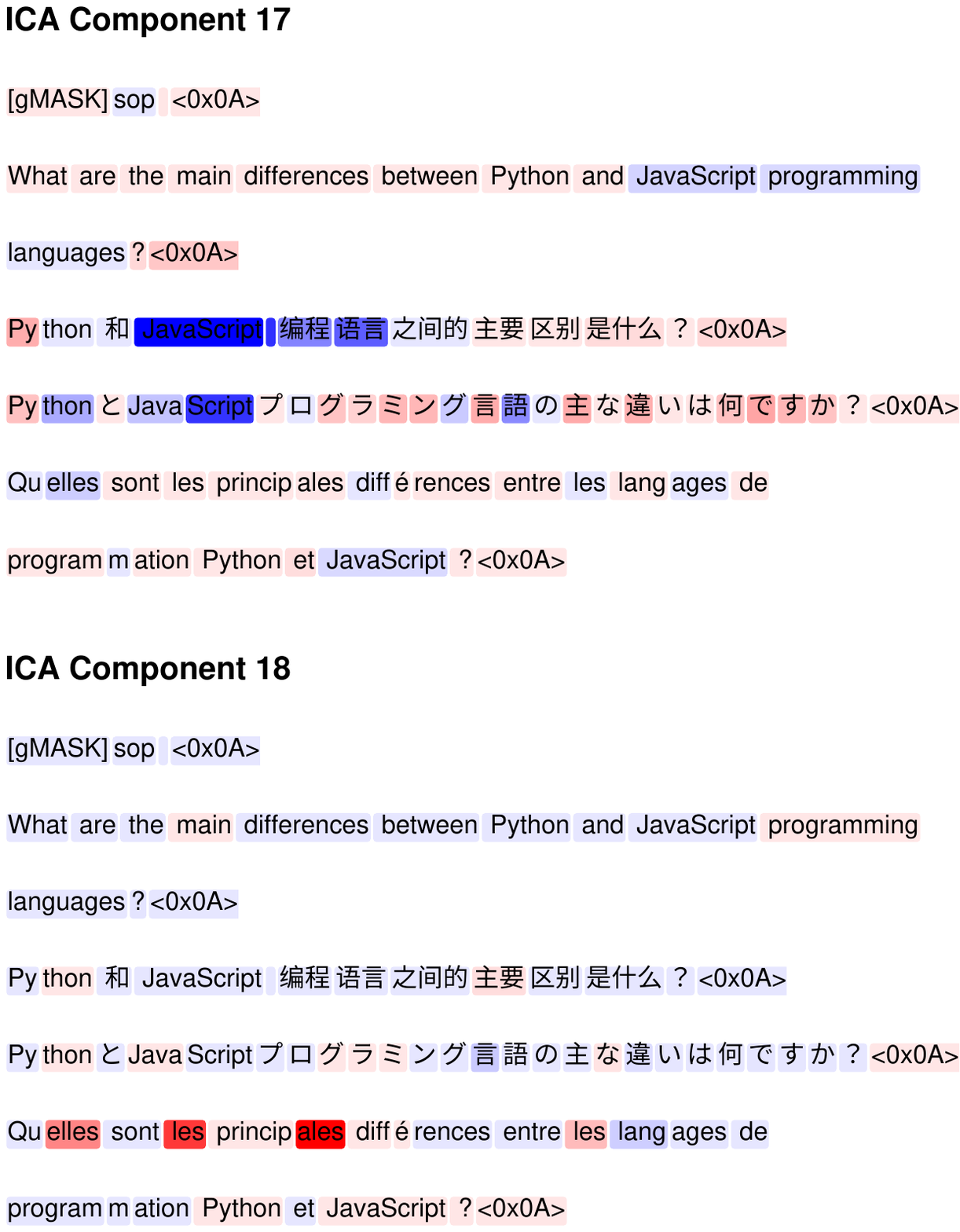}}
\caption{Interpreting ICA components by projecting their source signals back onto input tokens as attribution weights. High-weight tokens are visualized to reveal what linguistic features each functional network responds to.}
\label{fig::fbn_interpret_9}
\end{center}
\end{figure}

\begin{figure}[ht]
\begin{center}
\centerline{\includegraphics[width=\columnwidth]{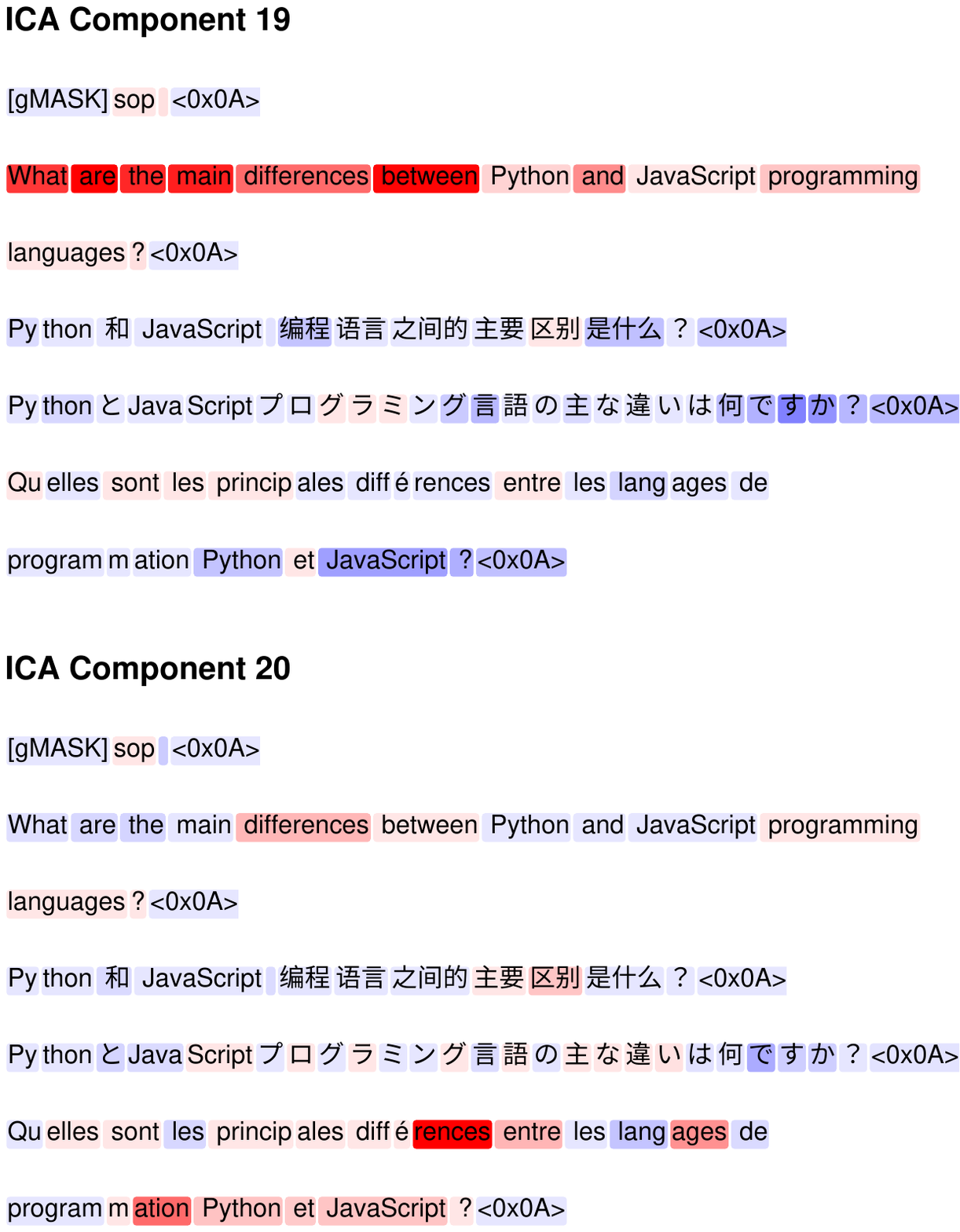}}
\caption{Interpreting ICA components by projecting their source signals back onto input tokens as attribution weights. High-weight tokens are visualized to reveal what linguistic features each functional network responds to.}
\label{fig::fbn_interpret_10}
\end{center}
\end{figure}

Figure~\ref{fig::fbn_interpret_1}, Figure~\ref{fig::fbn_interpret_2}, Figure~\ref{fig::fbn_interpret_3}, Figure~\ref{fig::fbn_interpret_4}, Figure~\ref{fig::fbn_interpret_5}, Figure~\ref{fig::fbn_interpret_6}, Figure~\ref{fig::fbn_interpret_7}, Figure~\ref{fig::fbn_interpret_8}, Figure~\ref{fig::fbn_interpret_9} and Figure~\ref{fig::fbn_interpret_10} show the example interpret results.

\end{document}